\documentclass[pdftex,12pt,a4paper]{article}
\usepackage{amsmath,amssymb,listings,subfig,hyperref,graphicx,tabularx,ragged2e,booktabs,caption}
\usepackage[acronym]{glossaries} % make a separate list of acronyms
\newcommand{\HRule}{\rule{\linewidth}{0.5mm}}
\makeglossaries

\newacronym{lapps}{LAPPS}{Location Aware Password Protection System}

\newacronym{atm}{ATM}{Automated Teller Machine}

\newacronym{srid}{SRID}{Spatial Reference System Identifier}

\newacronym{wgs}{WGS}{World Geodetic System}

\newacronym{gps}{GPS}{Global Positioning System}

\begin{document}
\begin{titlepage}

  \begin{center}
    \includegraphics[width=0.2\textwidth]{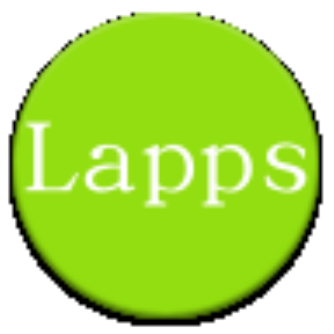}\\[1cm]

    \HRule \\[0.4cm]
    {\huge \bfseries Location Aware Password\\[0.2cm] Protection System}\\[0.4cm]
    \HRule \\[1.5cm]

    \textsc{\LARGE University Of Essex}\\[0.5cm]
    \textsc{\LARGE The Department of Computer\\[0.2cm] Science and Electronic Engineering}\\[1.5cm]

    \begin{minipage}{0.4\textwidth}
      \begin{flushleft} \large
        \emph{Author:}\\
        Chathura M. \textsc{Sarathchandra Magurawalage}
      \end{flushleft}
    \end{minipage}
    \begin{minipage}{0.4\textwidth}
      \begin{flushright} \large
        \emph{Supervisor:}\\
        Prof.~Kun \textsc{Yang}
      \end{flushright}
    \end{minipage}

    \vfill

  \end{center}
\end{titlepage}

\begin{abstract}
  \begin{itshape}
    This report describes the design and the implementation of a password
    protection system that has been proposed as an idea then developed
    by the author for his undergraduate final year project, at the University of Essex. 

    When designing the system the author has concentrated on adding
    extra layers of security to the traditional security systems
    without having to completely replace the existing security systems.

    \gls{lapps} is designed to strengthen the security of traditional password
    protection systems. This is achieved by adding several layers of
    protection to the passwords that most traditional password protection
    systems generate. The current implementation looks at the 
    Password/Pin numbers of Credit/Debit cards that are used on
    \gls{atm},though the underlying design of
    the system can be used in many  other scenarios. A password that
    is generated will be allocated to a particular user and to the \gls{atm}
    that is nearest to the user. \gls{lapps} ensures the following qualities
    of the passwords that it generates.

    \begin{itemize}
    \item Location Awareness

      The passwords are generated according to the user's geographical area,
      that they request their passwords from. So a password will only be
      active in just one location.

    \item Time Awareness:

      A password will only be valid for five minutes. The unused passwords
      will be discarded.

    \item  Dynamic

      The user has to have a new password each time he/she logs in. A
      password is generated to be used only once. 

    \item  User Oriented/Specific

      The received password can only be used by the requester, and can only
      be used on its allocated \gls{atm}.

    \item Two Factor Authenticity

      The confidential information will be secured using two-factor
      authentication.
    \end{itemize}
    For extra security, a Pin generating device has been introduced. This will
    produce an eight digit number that the user has to supply to the
    mobile application, before requesting for a password. The user
    can obtain a pin number by inserting his/her Debit/Credit card and the
    fixed password that has been allocated when the user registers
    with the system.
  \end{itshape}
\end{abstract}

%\section*{Acknowledgements}
%\thispagestyle{empty}
%I owe my deepest gratitude to my supervisor Prof.~Kun Yang for giving
%me guidance through out the processes of planning, designing, implementing and
%finishing my project. Then I am thankful for my family for providing
%financial means while I am at the University. It is my pleasure to thank
%Miss. Maria Ann Durbin for helping me by proof reading the documents.
%More importantly I would like to thank the University of Essex for
%giving me the knowledge and resources during my degree, because without it I
%will not be able to get this far. Lastly, I would like to offer my
%regards and blessing to everyone who has helped me in any way during this project.
%\clearpage

% Adds the table of content
\tableofcontents

\printglossaries

\section{Introduction}

With the increasing security issues in the modern password protection systems,
the security experts are concerned about their passwords more than
they did few years back. With the increasing possible computing power,
attackers are able to carry out multitudinous attacks on the passwords in a
small amount of time.(E.g. brute force attack).

With the advancements of the malicious attacks, it is almost hard to
trust the security of your confidential information that has been
secured only using long term fixed passwords. As a solution
\gls{lapps} adds varied layers of security to the existing password protection
systems. In section \ref{sec:traditional_vuls} the variations between the
layers and how they work will be explained.

\subsection{Background}
Mobile applications are rapidly growing with the escalating usage of
mobile devices such as hand held computers (Mobile Phones, Mobile
Pads). In the academia research on mobile software and hardware
technologies in the last decade have been increased. A core
fundamental element of "Pervasive computing" is to hide the underlying
mechanisms of systems to the user and perform actions with its
context-awareness. The idea is to sense and react to dynamic
environments and activities. Furthermore Location Information is a
compelling integral of context. So the Location aware computing is a
widely heard topic in this area. The authors of the article
\cite{hazas_scott_krumm_laccoa} write about the appliance of the
location awareness in "Invisible Computing". The papers
\cite{Marmasse00location-awareinformation} and \cite{1192785}
elaborates on "Location Awareness" for real world
applications. Moreover the latter suggests a "Pervasive Computing"
architecture that has been used on an implementation of a smart
environment that assists elderly persons to live longer. 

Location Awareness in most cases achieved by location
sensing. \gls{gps} has been widely used today for sensing geographical
locations using satellite signals. Despite the popularity of it, the
disadvantages and flows of \gls{gps} can not be neglected when
considering designing a "Location Aware" architecture. In the article
\cite{1203757} the authors analyse the reason why \gls{gps} is not the
optimal universal location sensing mechanism. The main reason
appeared to be that \gls{gps} does not work indoors, in particular in
steel-framed buildings. The resolution of it is a few meters and it may not
be adequate for some applications. Additionally the added weight, cost
and energy consumption might not be competent for some applications. Consequently academics in the "Lule ̊ University of Technology
" has designed an architecture \cite{Nord:2002:ALA:820747.821324} that
uses more than one sensing mechanisms to retrieve Location
Information, such as GPS, Bluetooth, WaveLAN, IR, HiperLAN and etc. Hence
that the availability, precision and accuracy of Location Information is
improved. Additionally they propose a Generic Positioning protocol (GPP)
for positioning information that exchange between devices and other
networked services.

Users may be apprehesive about the security when sharing their
location with a third party. The author sees this being a future problem while
carrying out further development of this project. One solution is to
control access to the information by letting user manage the delivery
and the accuracy of the location information using rule-based
policies. The paper \cite{1276918} introduces the "mix zone" model which
"anonymize" user's identity by restricting the position where users can
be located. Additionally the authors of the paper \cite{4343900} talk
about the privacy in Location aware computing.

\subsection{Related Work}
Considering the related work, the idea of a Location Aware Password
Protection system seem to be original. Albeit some work has been done
in the area of "Location aware access control" \cite{Cruz_alocation}
\cite{Michalakis03location-awareaccess} in combination of other contexts.  

\subsection{Vulnerabilities in conventional password protection
  systems}\label{sec:traditional_vuls}

The passwords that majority of password protection systems generate, are static
passwords. Meaning that they are assigned to the users in a fixed
manner. Hence these passwords can be used more than one time. Thus if
the password gets in to the hands of the unwanted, they will be able to
use it without the owner knowing nothing about it, in most cases. Or a
third party can use the password until the password would be changed
by the owner. But in most password protection systems if the password
is in the hands of the unwanted then they have no restrictions to change the
password without having the owners permission.

Malicious attackers are able to steal passwords from users by using advanced
techniques and equipments \cite{klein_pass_security_survery} (E.g
Skimming, password hacking). These stolen passwords can be reused to
get access to the corresponding user accounts. If bank card details
have been stolen using skimming devices, the card details can be
reprinted in to dummy \gls{atm} cards, and then the recorded password
can be used with its corresponding \gls{atm} card to steal money out of users accounts.

Attackers may be able to crack passwords using advanced algorithms
with high computational power. (The use of 'birthday attack' to
crack password hashes) \cite{klein_pass_security_survery}

Most password protection systems use One-Factor authentication. Which is
"Something a user knows" (E.g Systems that depend on security of just
one password). This is less secure. If the master password (one and
only security factor) is going to be compromised then the system will be open to any kind of malicious access.

\subsection{Objectives}\label{sec:objectives}

\begin{itemize}
\item Make the passwords dynamic so that each time a user logs in to
  the system he/she will use a new password.
\item Restrain the geographical locations that the passwords can be
  used, so that passwords can only be used within the users locations.
\item Make the passwords not reusable. So that used passwords will be inoperative.
\item Restrict the password's active time. Subsequently a user will
  have a short time frame to use the password. After the active time
  is up, the password will be invalid. 
\item Harden the security of systems by introducing a second authentication
  factor.
\item Create a uniform architecture that will accommodate above
  factors.
\item Implement the architecture as a solution to a real-world
  problem.
\item Evaluate the success of the architecture.
\end{itemize}

\section{LAPPS Architecture}\label{sec:lapps_arch}
The \gls{lapps} architecture follows the idea of having layers of protection
that can be variable. Which means depending on the desired level of
protection, the layers can be added or removed without having to
worry about the relationships between the layers, since the layers
does not depend on each other.

\gls{lapps} layers can be added to the existing password protection systems,
without having to entirely replace the old systems.

\begin{figure}[htp]
  \begin{center}
    \includegraphics[scale=0.3]{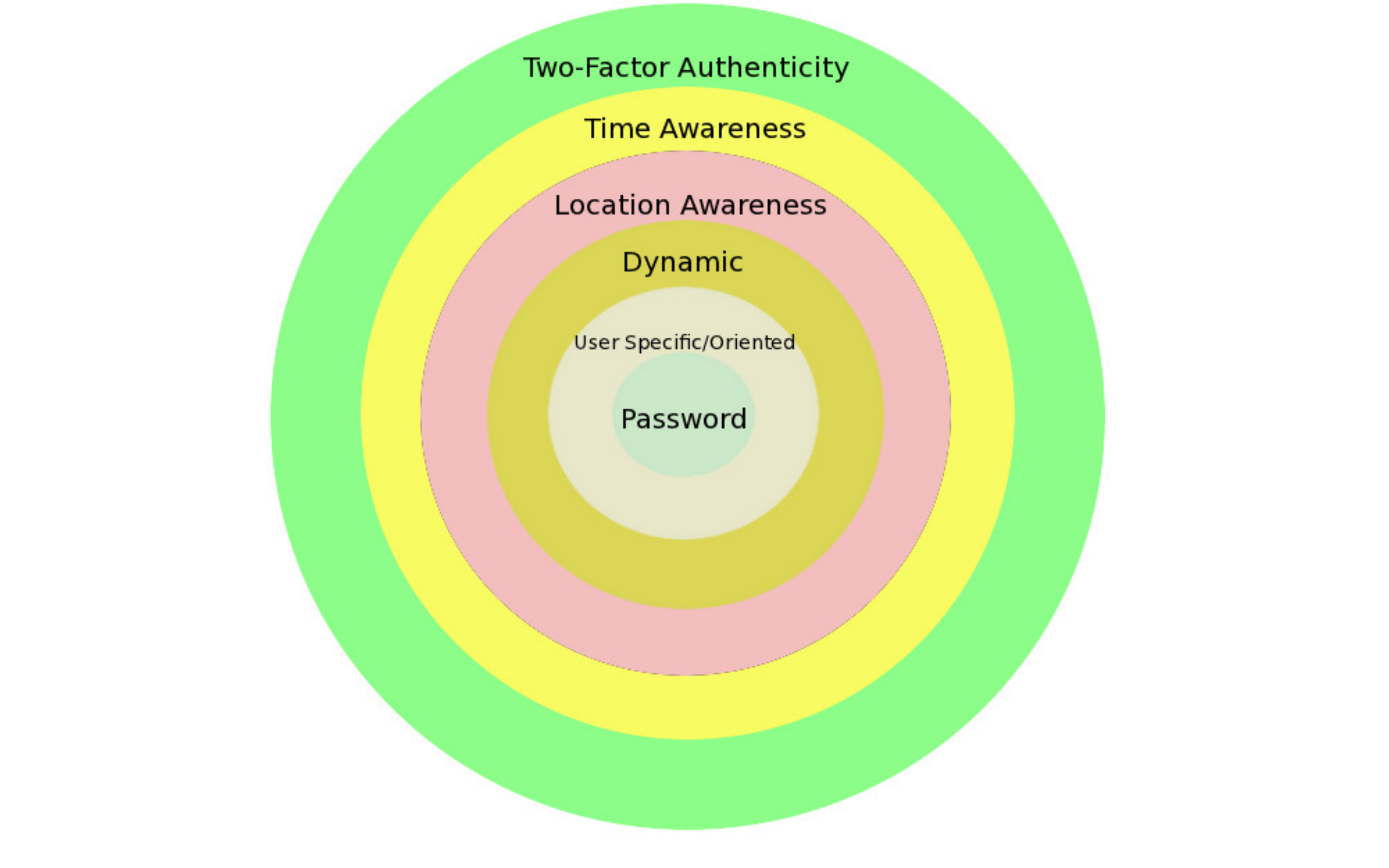}
  \end{center}
  \caption{The layers of LAPPS Architecture\label{fig:layers}}
\end{figure}

As shown in figure \ref{fig:layers}, the layers of \gls{lapps} wraps
around the password (base).

The following sections will elaborate on the core and the layers of the \gls{lapps} architecture.

\subsection{Password}
The core of the \gls{lapps} architecture is a standard password. A
character $p_i$ of a password $P$ can be $p_i \in L$, $p_i \in A$ or
$p_i \in S$ where $L$, $A$ and $S$ imply all possible letters,
Alphanumerical values and Symbols respectively.

The other layers that are wrapped around this string, will strengthen the security of the password.

\subsection{User Specific/Oriented}

This layer of \gls{lapps} operates on the database level of the system
called "lappStore". To make the password user specific, a password
$p_i$ is allocated to just one particular individual. A registered
user $U_i$ may be allocated to more than one password over time, but
he/she will only have one active password at a time. $HP$ denotes the
allocated user-password pairs in the history and $AP$ implies the active user-password pairs at present.\\

Such that:
\[
HP = \{U_1:p_1, U_1:p_2, U_1:p_3, U_2:p_5\}
\]
\[
AP = \{U_1:p_7, U_2:p_8, U_3:p_9\}
\]

If a user requests for a password one after another before the
former one expires, the former password will be deactivated and a new
password will be allocated. This avoids a user having more than one
password at a time. This ensures that there is only one active
password per user at a time, and only the owner can use it.

\subsection{Dynamic}
With \gls{lapps} the user has to request for a password each time they log
in. So that one password can only be used once. If a criminal gets
hold of a password illegally, then there won't be any use of the
password, if the owner of the password has already used it, or he/she
has requested for a new one.

In the current \gls{lapps} implementation a user is able to request
for a password using an Android mobile application called
"LAPPSMobile". When the client requests for a password, the server
will reply back with a new password only if the user's given
information is correct. The information included is the user's ID,
the registration number of the application, location information and
the eight digit number that has been generated by the Pin generating
device. The connection between the sever and the client is encrypted.

To be able to gain a registration ID for the mobile application, the
owner of the mobile phone and the particular instance of the
application has to be registered with the "LAPPSserver" (The server
side application of the \gls{lapps} architecture) along with the owner's user information.

The \ref{sec:location_aware},
\ref{sec:time_aware},\ref{sec:two_factor} sections will explain this extra information that has been sent to the server by the mobile application.
\subsection{Location Awareness}\label{sec:location_aware}

The location awareness is the main layer of \gls{lapps}. The reason being, this is the main factor that makes \gls{lapps} special. This particular
layer has been influenced to name the \gls{lapps} architecture. 

This layer makes the passwords active only in a particular geographical
area. When a user requests for a password, the user has to be
approximately in a password \emph{active zone}. If the user is not in a
password \emph{active zone} then the \gls{lapps} server may not
generate a password for that particular individual. An error message would
be passed back instead. The area size of the \emph{active zone} can easily
be variable according to specific requirements. If the user has
successfully received a password then to use the password he/she
has to be geographically present in the particular area that the received password is active. One password will be active only in one geographical area.

Because of this peculiar reason stolen unused passwords will be useless in
other geographical areas other than the \emph{active zone} of the particular
password. Although an attacker might be determined to use the password
in the \emph{active zone}, the other layers of \gls{lapps} make it harder for it to work.

\subsection{Time Awareness}\label{sec:time_aware}

The passwords that are generated by \gls{lapps}, are only active within a
limited time window. If the password is not being used within this time
frame, the password will be extinct. Depending on the requirements,
the time frame can be increased or decreased.

This leaves unused stolen passwords a limited amount of time to be
active. Hence the stolen unused passwords have to be used within a limited
amount of time and also within the exact corresponding password \emph{active zone}.
This makes stolen passwords nonetheless useless.

\subsection{Two-factor authenticity}\label{sec:two_factor}
\gls{lapps} uses two factor authentication to harden the existing layers of
protection. This layer can be removed easily if not desired.

Two-factor authentication uses two of the three well known authenticating
factors \cite{allen_pickup_two-factor_2007}. These two factors are:

\begin{itemize}
\item Something that user knows. (E.g: "A fixed password")
\item Something that user has. (E.g: "An \gls{atm} card")
\end{itemize}

A fixed password will be allocated to every user who is registered
with \gls{lapps}. This password can be used for this exact purpose
only. By using these two factors, the \gls{lapps} authenticates the user,
even before he/she gains a password. Only if the user is able
to authenticate using these two factors, he/she will be able to
gain a password. Any other way the server will bounce back with an
error message. Similar implementations have been introduced by other
organisations such as PinSentry device of Barclays Bank PLC \cite{AliJahaJaha2009rl}.

The most recent implementation of \gls{lapps} uses a fixed password and \gls{atm} card
information to generate eight digit number that will only be valid for
a very short period of time. This is attained by using a hash function
with salts. Section \ref{sec:lapps_atm_impl} explains how this is implemented.

\subsection{Evaluation on LAPPS Layers}\label{lapps_eval}

There are five layers in the \gls{lapps} architecture in total. The layers
can be removed or added to the stack of layers, depending on the
level of security desired by the user. Though new layers can be
added on top of these five layers.

\begin{figure}[htp]
  \begin{center}
    \includegraphics[scale=0.19]{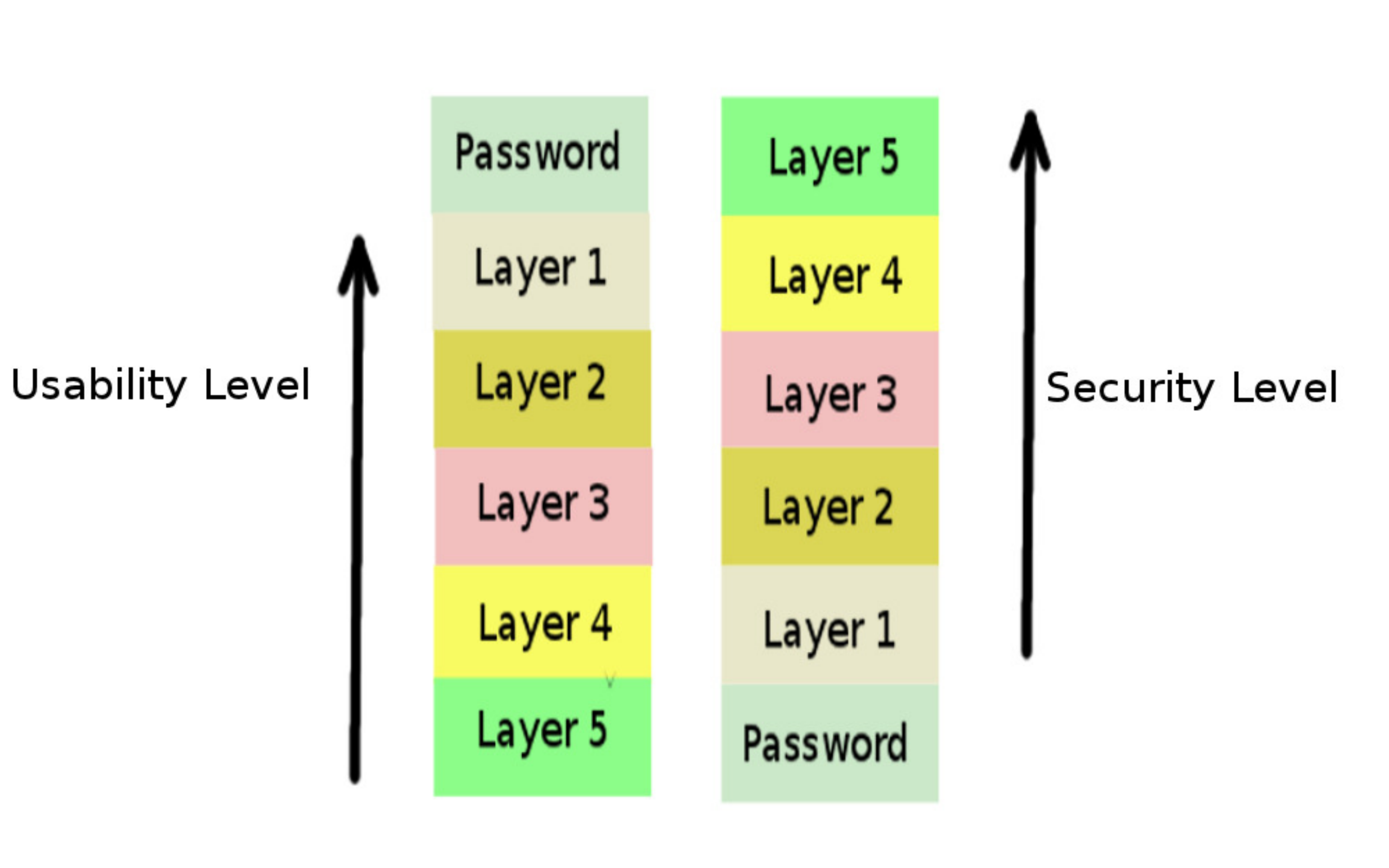}
  \end{center}
  \caption{Usability vs Security}
  \label{fig:layer2}
\end{figure}

Although the idea of adding more security layers is quite promising from
the security point of view, the more you add layers in to the stack
the less user friendly the system would be (Figure
\ref{fig:layer2}). So the security administrators have to find the
right security levels that would fit in to their needs. As an example
lets look at a system with and without the final layer the "Two-factor
authenticity". When this layer is present, as the first step the user
has to use their fixed password number and \gls{atm} card for example, to
obtain a second eight digit pin number. This number is then used to get
the final pin number that can be used to access the protected
information or service. This can be a tedious process granting that
the user is in a hurry or conceding that the user has not got the \gls{atm}
card with them when they need to get access to the particular service or data. 

An example for an extra layer of security would be to use the third
authentication factor, which is "Something that user is"
(E.g. fingerprints), with the other two authentication factors that is mentioned in section \ref{sec:two_factor}.

\subsection{LAPPS solution to the vulnerabilities in traditional
  password protection systems}

In section \ref{sec:traditional_vuls} the author has mentioned the
vulnerabilities that is in traditional password protection
systems. The following pointers will summarise the factors about the
solution \gls{lapps} architecture that has been explained in detail in section \ref{sec:lapps_arch}.

\begin{itemize}
\item A password that is generated by the \gls{lapps} is allocated to
  just one individual. 
\item There will only be one active password per user. So only the
  owner owns an active password.
\item The passwords are generated in a dynamic fashion so that a user
  may use a new password each time he/she access the secured
  data/service.
\item One password can only be used once. So that the stolen used passwords
  are dysfunctional.
\item The passwords will only be generated if the user can
  successfully authenticate. If a criminal steals a password
  requesting device (E.g. Smart Mobile phone with a registered
  "LAPPSMobile" Application installed) he/she will not be able to gain a
  password without supplying other information to the device.
\item The users have to be in a "Password \emph{Active zone}" to be able to
  request for passwords.
\item Every password is allocated to only one "Password Active
  Zone". Stolen unused passwords can not be used in multiple locations.
\item A password can only be used within its allocated "Password
  \emph{Active Zone}". Stolen unused passwords can not be used outside the allocated
  geographical area.
\item A password is only valid for a very limited amount of time. So
  that the user will have to use the password within this time
  frame. If the password is not being used within this time window,
  the password will be extinct.
\item To be able to authenticate, the user has to acquire an eight digit pin
  number using their fixed password and the corresponding smart card,
  for an example. So that stolen password requesting devices will be
  useless to a person who does not know the fixed password and does
  not own a Pin generating device with the corresponding fixed password.
\end{itemize}

\section{LAPPS ATM Implementation}\label{sec:lapps_atm_impl}

The current implementation of \gls{lapps} architecture focuses on hardening
\gls{atm} password systems by utilising its layers to fit in to the
conventional \gls{atm} systems.\\

Factors about traditional \gls{atm} password protection systems.
\begin{itemize}
\item A user may be assigned to a fixed password with its dedicated
  Credit/Debit card number.
\item The password can be used from anywhere in
  the world, along with its Credit/Debit card. The geographical
  location has no effect on the authenticity of the user.
\item The vulnerabilities mentioned in section
  \ref{sec:traditional_vuls} are valid in this case.
\end{itemize} 

There have been reported many crimes that are related to the security
of \glspl{atm} and other password protection systems
\cite{klein_pass_security_survery}. Such as, robberies, stealing
confidential information using hidden cameras and card readers and
hacking, with the advancements of technology.

\subsection{The Architecture of the LAPPS implementation for ATM
  systems}

\begin{figure}[htp]
  \begin{center}
    \includegraphics[scale=0.3]{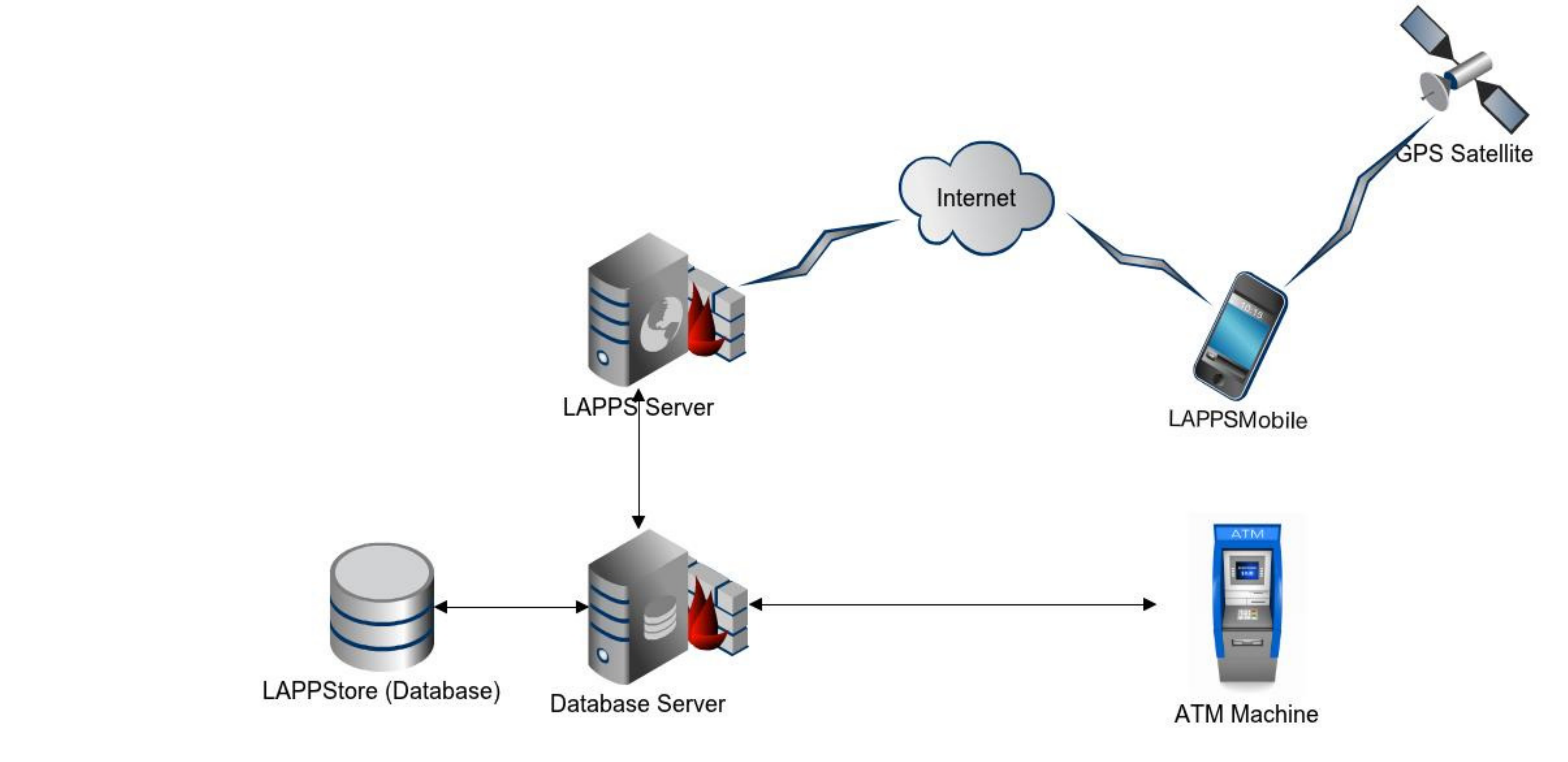}
  \end{center}
  \caption{The Architecture of LAPPS for ATM systems}
  \label{fig:lapps-arch}
\end{figure}

Figure \ref{fig:lapps-arch} shows, the components that are being
used in the \gls{lapps}. The server side application (LAPPSServer), client
application and Pin generating device are the components of this system. Section \ref{sec:layer_mapping} elaborates in what respect \gls{lapps}
uses these components in its various layers of security.

To be able to use the system the user has to have a smart phone that
runs Android platform.

The components of the system:

\begin{itemize}
\item "lappStore" is the main database of the \gls{lapps} system. The allocated
  passwords and user information are stored here. 
\item LAPPSMobile is the client application that the user can request
  passwords from. This application is developed on the Android
  platform and has to have the GPS functionality.
\item The LAPPSServer application replies to the requests from LAPPSMobile
  application, only if the authentication information is correct. 
\end{itemize}

\subsubsection{How a user would use "LAPPS for ATMs"?}

\begin{itemize}
\item User has to generate an 8 digit number from the Pin generating device
  by using his/her fixed password and the Credit/Debit card.
\item Then by staying no more than 20 meters away from the desired \gls{atm}
  that user intends to use, he/she can request for a password from
  the LAPPSMobile. And also may be asked to insert the 8 digit pin
  before the application sends a request to the server.
\item If the authentication details are correct then the server may
  reply with a password and the ID of the \gls{atm} that the password is
  allocated to. Else ways it will send back an error message.
\item Finally the user may log in to the \gls{atm} that the password is
  active on, using the password and his/hers \gls{atm} card.
\end{itemize}

\subsection{LAPPS Layer mapping to the "LAPPS for ATM" implementation}\label{sec:layer_mapping}

The following pointers will construe how the security layers of \gls{lapps}
have been implemented in to "\gls{lapps} for \gls{atm}" implementation.

\subsubsection{User specific/Oriented}
In a table called "allocation" in the lappStore, the LAPPSServer
stores the generated passwords with its allocated user and the \gls{atm}
machine ID. So that the password will only be allocated to one
user.

\subsubsection{Dynamic}
The passwords are not predefined. It is generated on demand as clients
request. If LAPPSServer has authenticated the user, then the password
will be generated.

\subsubsection{Location Awareness}

To be able to request for a password a user has to be less than 20
meters away from more than one \gls{atm}. The password is allocated to the
nearest \gls{atm} to the user only if the user is no far than 20 meters away
from it. And the password can only be used on the allocated \gls{atm}. 

\subsubsection{Time Awareness}

A password is only available for just 5 minutes after it has been
generated. If the user has not used the password on its allocated \gls{atm},
it will be expired.

\subsubsection{Two-Factor Authentication}

Before a user requests for a password, they will be asked to enter an 8
digit number. This has to be obtained from a Pin generating device by using
the user's \gls{atm} card and their fixed password.

\subsection{The components of "LAPPS for ATM"}\label{sec:components}

The configuration information of all of the following components have
been stored in properties files, in ".txt" format. On that account if the
information changes in the future the user is able to reconfigure the
software without having to change the code.\\

Such files may contain.
\begin{enumerate}
\item The URL of the database.
\item The name of the database.
\item The URL of the server.
\item The PORT number of the database management system.
\item The PORT number that the server is passively waiting/listening
  for connections from the clients.
\item The user names and the passwords for the database with different
  access permissions. 
\end{enumerate}

\subsubsection{The lappStore (Database)}

The database has been implemented on "PostgreSQL"
\cite{postgresql_documentation} database management
system. The reason for using this particular piece of software is that
it is open-source and the powerful PostGIS \cite{postgis_manual}
spatial database extension for PostgreSQL. This extension follows
Simple Feature Access for, SQL specification from Open Geospatial Consortium (OGC).\\
The database contains four tables. They are:

\begin{itemize}
\item user:\\
  This table stores the information about users in the system. A
  row of this table contains the user ID, registration ID, name,
  hashed fixed password. Passwords are hashed using SHA2-512 algorithm.
\item password:\\
  The hashed passwords are stored in this table with their expiry time.
  Passwords are hashed using SHA2-512 algorithm.
\item atm:\\
  This table contains the IDs of \glspl{atm} and their location
  information. This table contains all of the \glspl{atm} that the users can
  get access to.

\item allocation:\\
  An allocation contains a password, a user, an \gls{atm} ID and a Boolean flag
  that can toggle to be true if the password has been used, otherwise it
  is false (default). Since a user is allowed to use a password once, if
  this value is 'true' then the password is unusable. There will be one
  row in this table for every allocation (For sometime).
\end{itemize}

This database has been designed so that a minimum amount of data will be
stored, to keep the database less ponderous. To keep the database
clean, there is are triggers devised into the database.

The triggers trigger every time a user inserts a record in to the
allocation table. One trigger deletes the allocation rows that the
expiry time of the passwords have been exceeded and the other deletes
the rows that their password allocation has been already used. The
expired and used information in the 'allocation' table is useless
because they will not be used by the LAPPSServer in anyway after they
had its use. Although the passwords still will be stored in the
'password' table, because they are used by LAPPSServer to generate unique passwords.

The idea is that to keep this table clean as possible so that the
transactions can be performed somewhat faster.

\begin{figure}[htp]
  \begin{center}
    \includegraphics[scale=0.4]{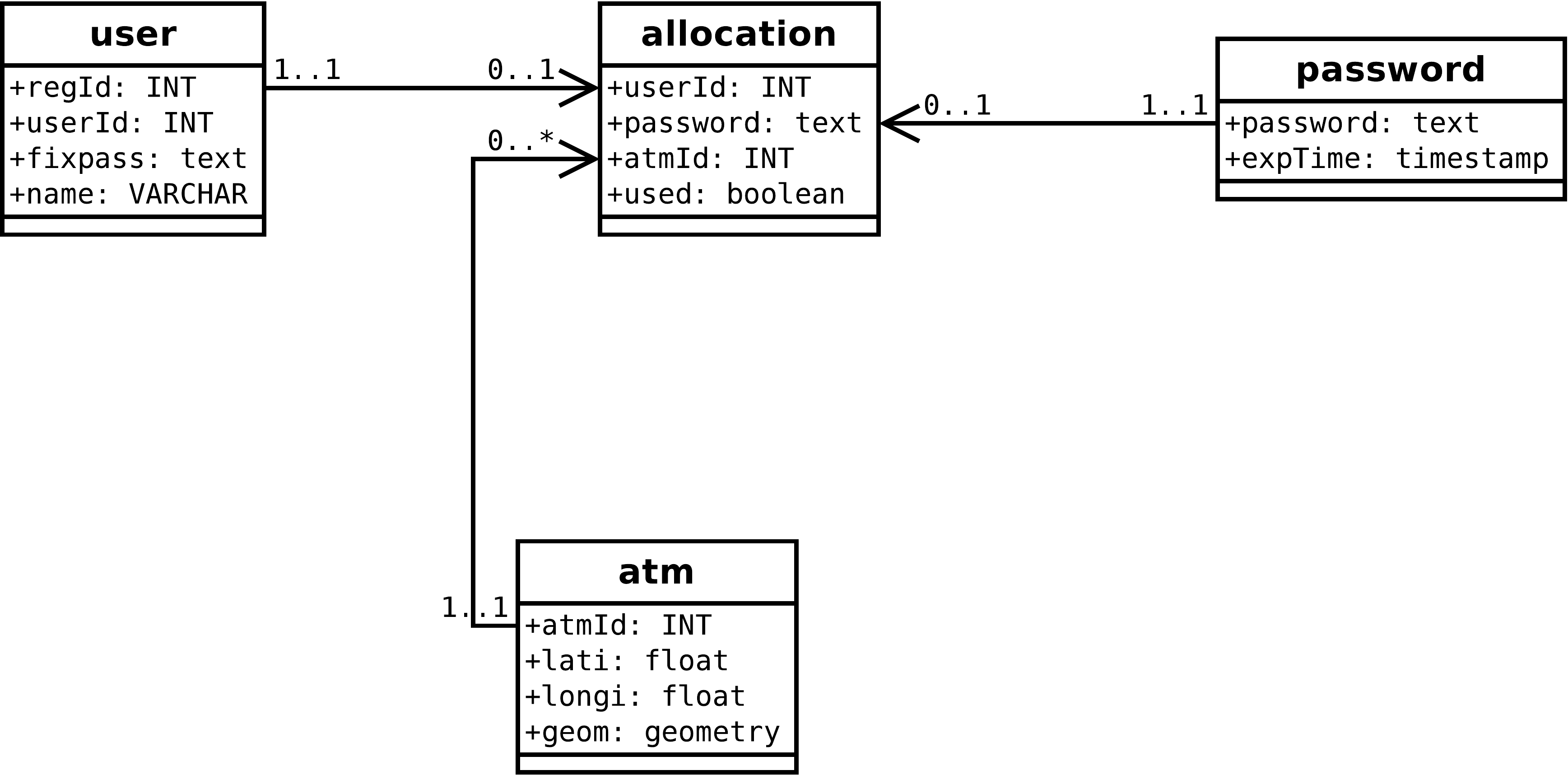}
  \end{center}
  \caption{The database design \label{fig:database_design}}
\end{figure}

\paragraph{Database design}\label{sec:database_design}
:\\\\
Figure \ref{fig:database_design} shows a diagram of the design of lappStore database.

The database contains four tables. Namely they are user, allocation,
password and atm. The relationships between them have been shown using
the black lines that connects the class boxes. The classes that have
been connected by a line with an arrow at the end are the foreign keys
that refer to the corresponding columns at the other end of the
relationship/line. For example "password" column in the "allocation"
table is a foreign key of the "password" column in the "password"
table, and so on. Hence the type of the foreign key column and the
corresponding column are the same.

The multiplicity factors of the tables shows in figure
\ref{fig:database_design} that a user can only have one allocation at
a time (0..1) but an allocation will always have a user (1..1). A
password can only have one allocation (0..1) but an allocation will
always have a password (1..1). An allocation will always have an ATM
(1..1) but an ATM may have zero or many allocations (0..*).

\subparagraph{user Table}\label{sec:user_table}
:\\

As the name suggests, the user table contains the information that is
relevant to the users of the LAPPS system. It is assumed that all of
these attributes of the users have been allocated and filled by the
bank. \\

Column description:

\begin{itemize}
\item regId   - The registered ID of the user's mobile application.
\item userId  - The ID of the user (Primary Key).
\item fixpass - The fixed password that has been allocated by the bank
  to the user, when they sign up with the bank.
\item name - The name of the individual. 
\end{itemize}

The current implementation allows users to have one application
registered per person. It is assumed that the users have already
registered their applications when they sign up with the bank. The
regId is stored in the users LAPPSMobile application instances.

\subparagraph{atm Table}

All of the ATM machines that belong to the system has to be
in this table. There should be one record per each ATM that
is registered with the system. The LAPPSServer uses the information in
this table to calculate the distance between the users and the
\glspl{atm}.

Column description:

\begin{itemize}
\item atmId - The unique ID of the ATM (Primary Key)
\item lati - The latitude geographical information of the ATM.
\item longi - The longitude geographical information of the ATM.
\item geom - The geographical information of type "\gls{srid}" 27700.
\end{itemize}

To create this special 'geom' column a special command had to be
used from the PostGIS library \cite{postgis_manual}. As parameters of
this function you can parse the "schema name", "table name",  "column name",
"SRID", type, dimension etc.\\

How this geometry column was added to the 'atm' table as follows.

\begin{verbatim}
SELECT AddGeometryColumn('public', 'atm', 'geom', 27700, 'POINT', 2);
\end{verbatim}

To calculate the column values of type \gls{srid} 27700 the author has
used the following command that iterates through all of the rows in
the 'atm' table and updates the 'geom' column.

\begin{verbatim}
UPDATE atm SET geom=PointFromText('POINT('|| lati||' '|| longi || ')', 27700);
\end{verbatim}

The 'PointFromText' function converts the received SRID 4326 (lati,
longi) value to SRID 27700 type value and stores it in the 'geom' column.

More information on spatial distance calculation will be discussed in the section \ref{sec:calc_atm_dist}.

Other parts of this database are triggers that are in service to
keep the database less bulky. These triggers fire after every time a record has been added to the table.\\

Trigger 1: Deletes the rows that are expired by time.\\

The pseudo code of the triggers:
\begin{verbatim}
BEGIN
DELETE FROM allocation as a USING password AS  pass WHERE
a.password=pass.password AND pass.expTime <= CURRENT_TIMESTAMP; 
RETURN NULL;
END
\end{verbatim}

Assume that the $\{a_i, \dots, a_n\}$ are $n$ allocations. Each
password $P_i$ has got its expiry time $ET_{P_i}$ associated with
it. The trigger function iterates through all $n$ allocation records
$A$ and deletes the records that are $ET_{P_i a_i} \leq CT$ where $CT$
is the current time. After the trigger has been fired we can say that,

\[
\forall a \in A : ET_{P_i a} > CT
\]

Trigger 2: Deletes the passwords that have been used.\\

The pseudo code of the triggers:
\begin{verbatim}
BEGIN
DELETE FROM allocation WHERE used=true;
RETURN NULL;
END
\end{verbatim}

If that the $\{a_i,\dots, a_n\}$ are $n$ allocations. The trigger iterates
through all $n$ allocations $A$ and deletes the allocation records that
the 'used' column of it $PT_{a_i}$ has been assigned to the value
\emph{true}. Furthermore, after the trigger has been fired we can say that,

\[
\forall a \in A : PT_a = false
\]

\subparagraph{password Table}\label{password_table}:\\

All the passwords that have been generated, are stored in this table
with its corresponding expiry date and time. \\

Column description:

\begin{itemize}
\item password - The password (Primary Key).
\item expTime  - The expiry time of the password.
\end{itemize}

The expiry time of the passwords have been calculated by the server. The
server adds 5 minutes to the current time and then stores the value in
the expTime column along with each generated password.

\subparagraph{allocation Table}\label{sec:allocation_table}:\\

This table contains all of the allocations that are alive on time. One
allocation has got the lifetime of 5 minutes. A trigger will delete
the expired rows. Ideally all of the passwords that are in this table,
are only unused allocations since the used allocations will be deleted
by one of the triggers. Of course there is a time window before the
triggers have been fired, where the used or expired allocations still
will be in this table.

Initially every time LAPPS has successfully generated a password, a
record will be added to this table. A user will have just one record
in this table at a time, since LAPPSServer makes sure that a user can have only one allocation/Password at a time.\\

Column description:

\begin{itemize}
\item userId - The ID of the user that the password has been allocated
  to(Foreign key).
\item password - The allocated password (Foreign key).
\item atmId - The ID of the ATM that the password has been
  allocated to (Foreign key).
\item used - Boolean flag; true if the password has been used, false otherwise.
\end{itemize}

\subsubsection{Pin Generating Device}

Users have to use this device to generate an 8 digit pin number before
they request for a dynamic password. A user may insert their
Credit/Debit card in to the Pin generating device and insert his/her
fixed password. Using the card and password an 8 digit password may be
generated. 

\begin{figure}[htp]
  \begin{center}
    \includegraphics[scale=0.5]{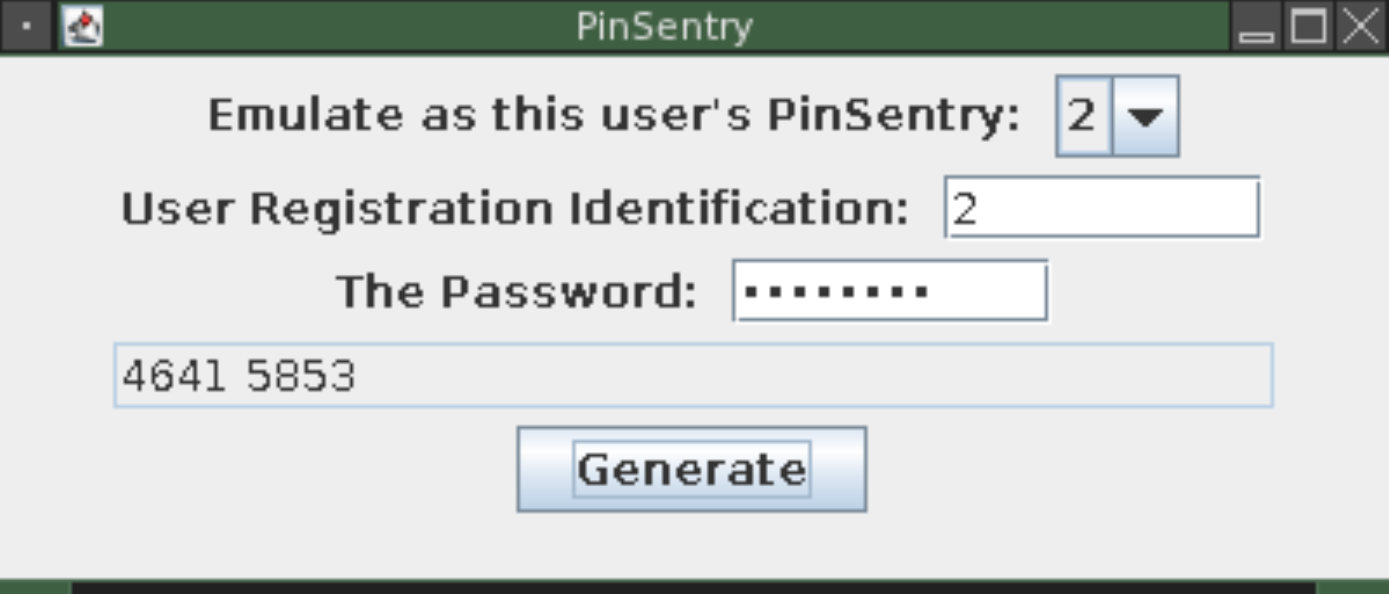}
  \end{center}
  \caption{Pin Generating device software \label{fig:pinsentry}}
\end{figure}

How a user would use the application?\\

\begin{enumerate}
\item Select the user's register ID from the drop down selection menu.
\item Enter the user's registration ID.
\item Enter the user's fixed password
\item Click the 'Generate' button.
\item Finally the generated password will be generated on the screen.
\end{enumerate}

Although the users enter all of their authentication information in
this application, a real Pin generating device, does not prompt the
user to insert this information. Instead this information apart from
the user's fixed password, is obtained from the user's Credit/Debit
card when it is being inserted into the device. For that reason only
for the testing and debugging purposes the above design has been
used. Besides the project's emphasis is on the algorithm that is being used to generate the pin numbers.

\paragraph{Algorithm}\label{sec:algorithm}
:\\\\
Firstly the following algorithm generates a hexadecimal value
$X$, assuming that the byte array that has been returned by $H$ the
SHA2-512 hashing algorithm, has been converted to hexadecimal. If all
possible time values are $TM$, given a time stamp of the current time
rounded to the nearest minute is $TT$ we can say that $TT\in
TM$. The $TT$ is rounded to the nearest minute since it removes added
complications to the algorithm by the "Seconds" fraction of the time
stamp, when the server authenticates the pin number (More on pin number
authentication in section \ref{sec:server_auth_alg}). We assume that $\{U_i,\dots,U_n\} $ are $n$ parties who are registered with
LAPPS. $ID_{U_i}$ denotes the $U_i$'s unique identity. $FP_{U_i}$ implies the fixed password of $U_i$.

\[
X = H( H(FP_{U_i}) + t + ID_{U_i})
\]

Secondly loop through the characters $x_i$ of $X$ and retrieve the
first four characters that are $x_i \in D$ where $D$ denotes digits
from 0 to 9. Reverse the hexadecimal string and then again carryout
the same procedure to find the final four digits of the eight digit number.

Every minute this pin number will change due to how hash functions
work with the salted time stamp. Which means that the pin numbers that
are being generated by using this technique will only be valid for around a minute or two (Further discussion is in section \ref{sec:server_auth_alg}).

It is highly unlikely that someone would be able to guess or crack
this password, using today's computational power within this limited
time frame. The added salts make it even harder.

\paragraph{Software Design}\label{sec:pinsentry_design}:\\

\begin{figure}[htp]
  \begin{center}
    \includegraphics[scale=0.43]{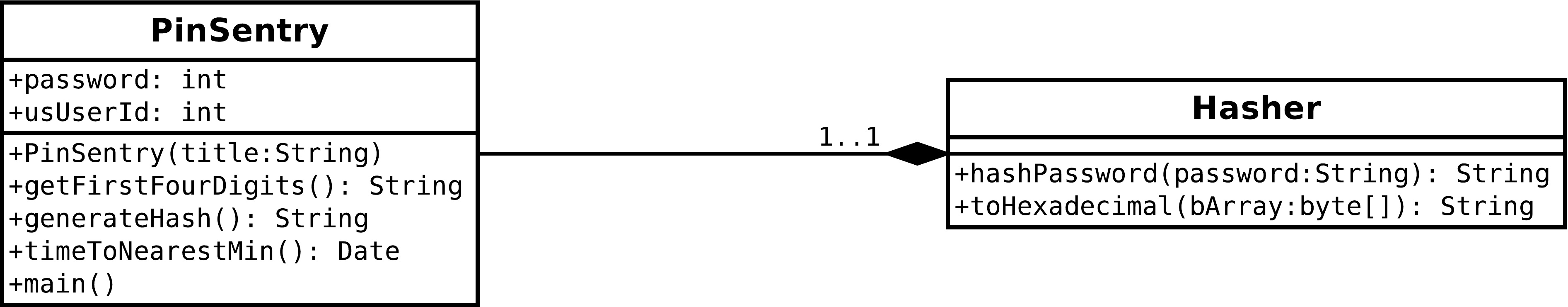}
  \end{center}
  \caption{The class diagram of the Pin Generating device software \label{fig:Diagram2_pinsentry}}
\end{figure}

The software consists of two classes Hasher and the PinSentry. Figure \ref{fig:Diagram2_pinsentry} shows a class diagram of these two classes.

The PinSentry class creates all the Graphical User Interface for the
users to insert their authentication information and to display the
generated pin numbers. The static methods in the Hasher class has been
used by the PinSentry class to generate the hash values and to convert
them to Hexadecimal values.

As shown in figure \ref{fig:Diagram2_pinsentry}, a Pinsentry has a Hasher (1..1).

Hasher uses Java's inbuilt libraries to hash the strings. In
particular the "java.security.MessageDigest" class. The PinSentry
class may only produce a pin number only if the given authentication
information were correct. Although do not forget that even if the
PinSentry device allowed the users to enter invalid fixed passwords, because
the inserted fixed password completely depends on the pin number
generation, the server will deny the pin numbers that are generated with invalid pin numbers.

\subsubsection{LAPPS Server}\label{sec:lapps_server}
The server application has been developed using the Java
technology. This Java daemon listens on a port in the host
computer until a client connects to it. The current implementation
only waits for "GETPASS" requests. Though more services can easily
be adapted to the server in future (More on section \ref{sec:client_server_communication}). When the client sends a "GETPASS"
request to the server with it's authentication information and
location information, the server daemon will reply with a password
only if the authentication information is correct. 

\paragraph{Software Design}\label{sec:lappsServer_software_design}:\\

\begin{figure}[htp]
  \begin{center}
    \includegraphics[scale=0.28]{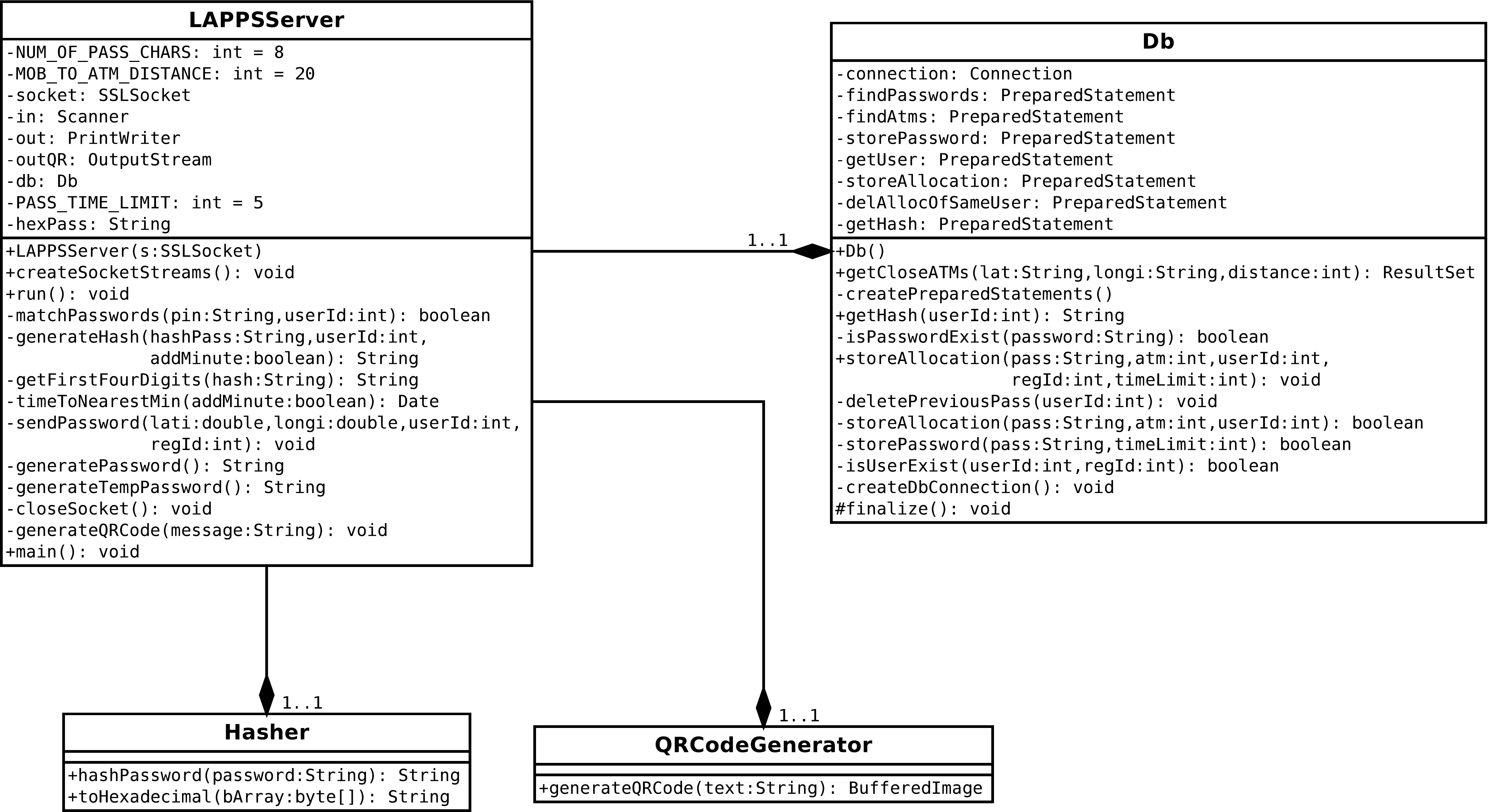}
  \end{center}
  \caption{The class diagram of the LAPPS server}
  \label{fig:lappsServer_software_design}
\end{figure}

Figure \ref{fig:lappsServer_software_design} shows a diagram of
design of the LAPPS Server.

LAPPS server is a daemon service that is preferably running on a
server machine. Which means that powerful hardware is needed to handle
the number of incoming requests from the clients. The current client
application is designed for the Android platform. The LAPPS server is
able to handle "GETPASS" requests from the clients as described in the
beginning of the \ref{sec:lapps_server} section. 

Figure \ref{fig:lappsServer_software_design} illustrates the
relationships between the classes. LAPPSServer class has an instance of the Db class, a Hasher and a QRCodeGenerator (1..1).

The most befitting way to elucidate the relationships of the bare mechanisms
inside the LAPPSServer is to analyse it's work flow. So the following
section \ref{sec:server_workflow} unravels the inside code
mechanisms of the LAPPSServer which responds to requests that it
receives (described in the protocol section \ref{sec:lapps_protocol}) by the client applications.

\subparagraph{The server's work flow}\label{sec:server_workflow}

\begin{enumerate}
\item The LAPPSServer waits passively for client requests.
\item The client connects to the server. Then the server will be
  passively waiting for requests from the clients.
\item Client application sends a "GETPASS" request to the server with
  other geographical and authentication information (More on password requests in section \ref{sec:client_server_communication}).
\item LAPPSServer validates the 8 digit number (More on 8 digit pin number
  generation in section \ref{sec:algorithm} and validation in section \ref{sec:server_auth_alg}).
\item If the 8 digit number does not match then an error message will
  be sent back to the client encoded in to a QR Code(More on error
  messages in section \ref{code:fail:error}).
\item If the 8 digit number is correct then, LAPPSServer checks if
  there are any Automated Teller Machines that is less than 20 metres
  away from the user. (More on finding the nearest ATMs in section \ref{sec:calc_atm_dist}
\item If there are no any \glspl{atm} then an error message is sent
  back to the client encoded in to a QR code (More on error messages in section
  \ref{code:fail:error}).
\item Then a random unique password is created that has not been
  generated before (More on password generation in
  section\ref{sec:password_generation})
\item The generated password is hashed using SHA2-512 one way hash
  function. The returned byte array from the function is then
  converted to Hexadecimal. This is done so this value will be stored
  in the database as a substitute to storing the password in plain
  text. Accordingly the security of the passwords strengthened.
\item If there is an allocation record that belongs to the same user
  already in the database, then the server deletes it before storing
  the new allocation into the table, as one user is allowed to have
  only one active allocation/password at a time (More on storing
  allocations in section \ref{sec:allocate_new_pass}).
\item The generated password and the closest \gls{atm} are allocated to the user by  storing a record in the 'allocation' table with the generated password, the ID of the \gls{atm}, the ID of the user, the registration ID of
  the user's application and the expiry time of the allocation (More on storing
  allocations in section \ref{sec:allocate_new_pass}). 
\item A response message is created with the ID of the allocated
  \gls{atm} and the password (More on successful messages in section \ref{sec:client_server_communication}).
\item Then this respond message is encoded in to a QR code, which in
  size vary according to the size of the message (More on QR code
  generation in section \ref{sec:qr_generation}).  
\item Finally the QR code with the generated password and the
  ID of the allocated \gls{atm} is sent back to the client.
\end{enumerate}

\paragraph{Allocating new passwords}\label{sec:allocate_new_pass}:\\

Let $\{a_i,\dots,a_n\}$ be all $n$ active allocations $A$ that are
in the database. $U_i$ is a user that owns an allocation. This user
may only have one and only allocation $a_{u_i}$ at a time in the
'allocation table'. Such that:

\[
\exists ! a \in A: a \mapsto U_i
\]

Consequently before allocating a new password the LAPPSServer queries
the database to ensure that there will not be another active allocation in the
database. If there is already a valid allocation, the sever deletes it
before storing the new allocation/password in to the database. How
this can be achieved is presented in the following pseudo code of a SQL query.

\begin{verbatim}
DELETE FROM allocation WHERE userId=<USER_ID>
\end{verbatim}

Then to store the new allocation the server executes an SQL
command. The following pseudo code shows how this is done.

\begin{verbatim}
INSERT INTO allocation(userId, password, atmId, used)
VALUES(<USER_ID>, <GENERATED_PASSWORD>, <ALLOCATED_ATM>, false)
\end{verbatim}

\paragraph{Communication between the server and client} \label{sec:client_server_communication}:\\

The communication between the client and the server is carried out by
using TLS over TCP/IP sockets. The reason for using TCP/IP sockets is
the low overhead when transferring data over. The TLS layer will
encrypt the data since the data being transferred are private and
confidential.\\

Java JDK comes with a tool called "keytool" that can be used to
generate key stores and to export them to certificates. In this
project the author has used the "keytool" to create the key
stores. This tool can be used to generate key stores for debugging
purposes. A commercial certificate that is signed by a certificate
authority (CA) can be created by buying certificates from a well known
certificate authority such as "Verisign" and importing/signing your certificates by the bought certificate.

After a secured connection has been established the client and the server
uses the following messaging techniques to communicate with each
other.

The client application sends "GETPASS" requests to the server to
request dynamic passwords to be able to use on \glspl{atm}. 

An example of a "GETPASS" request.\\

\emph{GETPASS $<$8 digit pin$>$ $<$userID$>$ $<$reg ID$>$
  $<$latitude$>$ $<$longitude$>$}\\

If the request is valid the server generates a password and finds
and designates this password to the nearest Automated Tailor Machine
to the user. Finally it creates a response message by concatenating
this information. \\

A response message of successful transaction looks like as follows.\\

\emph{ SUCCESS: $<$atm\_id$>$ $<$password$>$} \\

And a response/error message of unsuccessful transaction looks like as
follows.\\

\emph{ FAIL: $<$ The Error message $>$}\\ \label{code:fail:error}

The LAPPSServer has been coded so that new mechanisms can be easily
implemented in the future. It has been achieved by modularising the code
that deals with divergent request messages. Alternately the
encapsulated code makes the program code easy to read. \\

The following pseudo code shows how this is done in the LAPPSServer
class. To demonstrate how code that deals with new request codes can
be added, the following code contains a dummy block of code that deals
with "HELP" requests. 

\begin{verbatim}
while (true) {
  if (connection.hasRequests()) {
   
/****Deals with GETPASS requests***/ 
   if (connection.next().equals("GETPASS")) {
      recieveInformation();
      if (matchPasswords() && hasATMs()) {
        sendPassword();
      } else {
        sendErrorMessage();
      }
   }
/**********************************/ 

/****Deals with "HELP" requests***/ 
   if (connection.next().equals("HELP")) {
     //Add code to deal with "HELP" requests.
   }
/**********************************/ 

   // Code for handling other request codes can be added here
}
\end{verbatim}

\paragraph{The LAPPS Protocol}\label{sec:lapps_protocol}

:\\

\begin{figure}[htp]
  \begin{center}
    \includegraphics[scale=0.3]{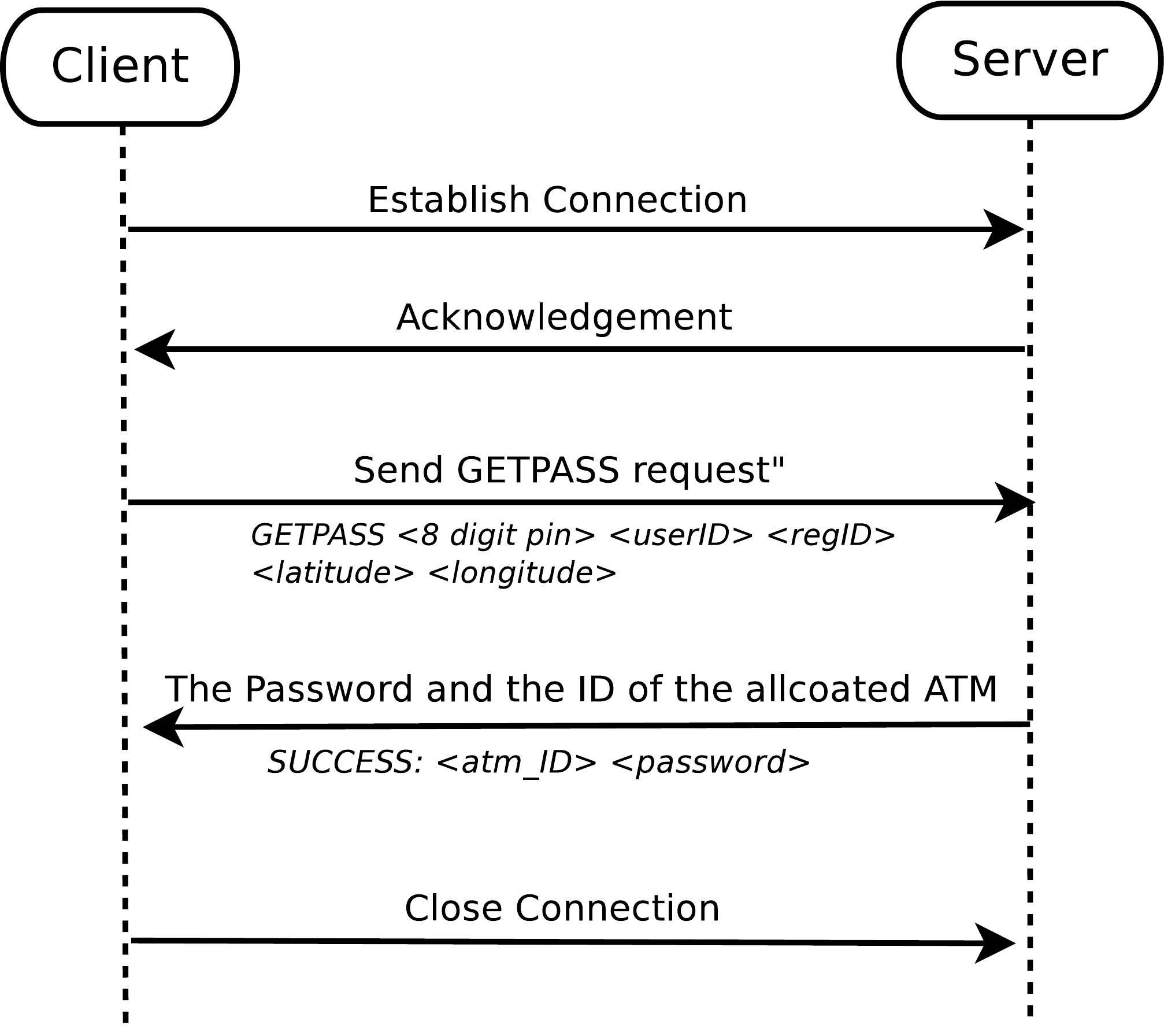}
  \end{center}
  \caption{The Protocol of the LAPPS Architecture}
  \label{fig:protocol}
\end{figure}

A successful transaction will undertake following steps to
successfully retrieve a password from the LAPPS Server as shown in
"figure \ref{fig:protocol}".

\begin{enumerate}
\item Client establishes a TLS over TCP/IP Connection to the server.
\item Server responds with an Acknowledgement message to the client.
\item Client sends a "GETPASS" request to the server with the
  additional information.
\item If the authentication information are correct then the server
  responds with a password with the ID of its allocated \gls{atm}.
\item Client closes the connection. 
\end{enumerate}

If the server is unable to send the acknowledgement message, the
client would not carry out any other requests. Since this message
ensures the client that it is open for any requests from the client.

\paragraph{Password generation}\label{sec:password_generation}
:\\
A character in a password is generated by;

\begin{enumerate}
\item Store the valid characters in an array. 
\item Then by using a random number generator, generate a random
  number that is no greater than the size of the array. 
\item Finally get the character that the subscript value of it is equal
  to the generated random number.
\item This procedure is carried out eight (The number of characters in
  the password) times.
\item Finally To make the generated password unique, the server
  queries the "password" column in the "password" table and look for
  pre-generated passwords. If the password already exists, then
  another password is generated and compared until it generates a
  password that does not exist in the database. Although this adds a
  high overhead on the password generation process as the database
  grows.
\end{enumerate}

The length of the password can easily be variable as easily as changing
an Integer variable in the LAPPSServer class.

The pseudo code of the SQL query that looks for existing matching
passwords in the 'password' column.

\begin{verbatim}
SELECT password FROM password WHERE password=<generated_password>
\end{verbatim}

If this query returns a Result Set then this password already exists in
the database.\\

More suggestions on password generation:\\

To make the sequence of characters even more random, the array can be shuffled
former to the selection. Although this will add an overhead when generating a password.

\paragraph{Finding the nearest Automated Tailor Machine}\label{sec:calc_atm_dist}

:\\

The author has used PostGIS extension to calculate the distances between two
points in spatial surfaces as it allows programmers to manipulate spatial objects in PostgreSQL databases. 

In the table called "atm" contains the location information of all of
the \glspl{atm} that belongs to the system. The location information are
stored as 'Geometry' typed instances. 

The geometrical information that is retrieved are likely to be in the
format of \gls{wgs} 1984 system. The \gls{srid} of this is '4326'. The
geometrical instances are stored in the type of \gls{srid} 27700 since
the author lives in England, United Kingdom. Because depending on where the
objects are geographically located the \gls{srid} changes.

The properties of the \gls{srid} 2770 in the EPSG Geodetic Parameter
Registry \cite{EPSG_web} is as below.

\begin{itemize}
\item Name            : OSGB 1936 / British National Grid \\
\item Code            : EPSG::27700 \\
\item Type            : ProjectedCRS \\
\item Area Description: United Kingdom (UK) - Great Britain - England and Wales onshore, Scotland onshore and Western Isles nearshore; Isle of Man onshore.\\
\end{itemize}

Having got these locations stored in the database, to find the nearest
\gls{atm} machine to the user the server follows the following
procedure. 

\begin{enumerate}
\item Convert the user's geographical information in to the \gls{srid}
  27700 format.
\item Query the database, and calculate the distance between the
  user's geographical spatial point and the \glspl{atm} points.
\item Then retrieve the \gls{atm} information of the \glspl{atm} that
  are in distance of less than 20 meters away from user. Therefore in the
  retrieved result set $\{r_i,\dots, r_n\}$ of all $n$ elements $R$
  the calculated distance is stored in the virtual column
  'distance'. Such that,
  \[
  \forall r \in R: distance_r < 20
  \] 
\item Order the Result set by the distance.
\item Get the top/first element of the Result Set.
\end{enumerate}

All of above steps can be done by using one query.

To calculate the distance the author has used the st\_distance\_sphere()
function of PostGIS \cite{postgis_manual}. 

The author used two variations of the query that had two widely
different performance levels.

The first query was a simple query that calculates the distance from
the given user's geographical location (latitude, longitude) and the
all other \glspl{atm} in the database. Then it orders the result by
the distance and finally retrieve the first row which is the closest
\gls{atm} to the user. The SQL query is as shown below.

\begin{verbatim}
SELECT atmId, st_distance_sphere(PointFromText('POINT(latitude longitude)',
27700), geom) AS Distance  
From atm 
order by distance 
limit 1;
\end{verbatim}

This is not the optimum solution. This query calculates the distance
for every \gls{atm} in the database which can easily be millions in real
life. So due to this issue there will be an unacceptable overhead when
executing this SQL query.

The SQL query that is currently implemented in LAPPSServer is the
following. 

\begin{verbatim}
SELECT atmId, st_distance_sphere(PointFromText('POINT(latitude 
longitude)',27700), geom) AS distance
FROM atm
WHERE geom && expand(PointFromText('POINT(latitude longitude)',
27700), 20) AND st_distance_sphere(PointFromText('POINT(latitude longitude)',
27700), geom) < 20 
order by distance
limit 1;
\end{verbatim}

This query creates a bounding box of 20 metres around the interest
point (User's location) and filters the rows that fits in this
box. Then it calculates the exact distances to the user location, orders the points by their distance to the user and finally
selects the first row/point. 

The author has found this query very efficient that it performs much
faster than the first query.

\paragraph{Authenticating users}
:\\
The server authenticates the users by checking the validity of the 8
digit pin number, the ID of the user and the registration number.

Let $ID^R_{U_i}$ be the received unique identification of the user $U_i$ and the
$ID_{u_i}$ is the identification of $U_i$ in the database, $PN^R_{u_i}$ is the
received 8 digit pin number that has been received by $U_i$ and
$PN_{U_i}$ is the calculated 8 digit number by the LAPPSServer (More
on calculating the 8 digit number in section
\ref{sec:algorithm}) if $RN^R_{U_i}$ is the received
registration number of $U_i$ and $RN_{U_i}$ the stored registration
number of $U_i$ in the database, we can say that only if $ID^R_{U_i} =
ID_{u_i}$ and $PN^R_{u_i} =  PN_{U_i}$ and $ RN^R_{U_i} =  RN_{U_i}$
that $U_i$ is authoritative.

The registration number is assigned to the particular instance of the
LAPPSMobile android mobile application. When $U_i$ registers
his/her mobile device with the user account, this field will be
allocated.

\subparagraph{Algorithm to authenticate 8 digit number}\label{sec:server_auth_alg}
:\\
To check if the 8 digit number is correct. The server uses the
same algorithm as the Pin generating device. The author has described how this
is done in section \ref{sec:algorithm}. Server is capable of
calculating the same algorithm independently since the user ID and the
corresponding fix password are stored in the lappStore database.

The server calculates the same algorithm $f$ using the current time stamp
rounded to the nearest minute. But when the server does this, the
rounded minute might be different to the one that the Pin generating
device has been used. For an instance, if the user $U_i$ has generated
a pin number $PN^R_{U_i}$ at "12.30:45" then server calculates the pin
number at "12.31:15", the rounded time stamp of the server will be
12.31:00 where as the Pin generating device has used "12.30:00" when
calculating the password. For that reason, if the generated 8 digit
number $PN_{u_i}$ of $U_i$ does not match it will generate another pin
number using $f$ but rounding up the time stamp $TT_{ms}$ after removing a
minute in milliseconds $ms$ from the time stamp. To make the following
notation simple and clear $GU$ denotes the authenticated users. Such that,

The first attempt to authenticate:
$$PN_{u_i} = f(TT_{ms})$$
$$If$$
\[
PN^R_{U_i} = PN_{U_i}
\]
$$Then$$
\[
U_i \in GU
\]

The second attempt to authenticate if the first attempt was unsuccessful:

$$If$$
\[
PN^R_{U_i} \neq PN_{U_i}
\]
$$Then$$
\[
PN_{u_i} = f(TT_{ms}-60000_{ms})
\]
$$\therefore$$
$$If$$
\[
PN_{u_i} = PN^R_{U_i}
\]
$$Then$$
\[U_i \in GU\]

Finally if $U_i$ is still $U_i \notin GU$ then the received password request is
not approved.\\

In other words, if the received pin number is correct it will match with one of the two pin numbers that is generated by salting two different time stamps.

\paragraph{QR Code Generation}\label{sec:qr_generation}:\\

QRCodeGenerator class uses the "com.google.zxing" library to create QR
codes the a given string. In particular an instance of QRCodeWriter
class has been used to perform the encoding of the
strings. Alternately the encode() method returns a QR code of type
'BufferedImage' with the given string encoded in it.

Conventionally there is a fixed size to the QR codes that applications
generate, that specified by the developer. But the speciality of
this piece of code is that it generates the QR codes that only fits
the string that is supplied. So it will not produce large QR codes
when only a fraction of it has been used. This is important because
small QR codes are beneficial when transferring them over networks.

A QR code with a "SUCCESS" message (section
\ref{sec:client_server_communication}) encoded inside is about the size
of a '29 x 29' pixel image. An average size of a QR code with a "FAIL" message
(section \ref{sec:client_server_communication}) coded inside is about
the size of a '33 x 33' pixel image. These images are very small so
that they can be transferred over the networks efficiently. Although
the performance of the system can be dramatically improved if the
messages are passed as text, the added QR code functionality makes it more enjoyable.

\subsubsection{LAPPSMobile}

This mobile application is the only way of requesting passwords from
the LAPPS Server. This application is designed for the mobile phones
that run Android platform. The application requires access to the
Internet and geographical location information. 

As described in section \ref{sec:lapps_protocol}, this application
sends "GETPASS" requests to the server with other authentication and
location information, to retrieve a password.

Additionally, this application is capable of understanding the message
codes that are sent by the server. Currently there are only two error
codes. they are "SUCCESS" and "FAIL". The author has elucidated these two
codes in section \ref{sec:lapps_server}.

If the application receives a message with a "SUCCESS" code, it will
print out a message as shown in figure \ref{fig:lapps_mobile_ss_c}.

The following figure \ref{fig:lapps_mobile_ss} shows the screen shots
of the completed LAPPSMobile Application.

\begin{figure}[htp]
  \begin{center}
    \subfloat[The Pin request button]{    \includegraphics[scale=0.228]{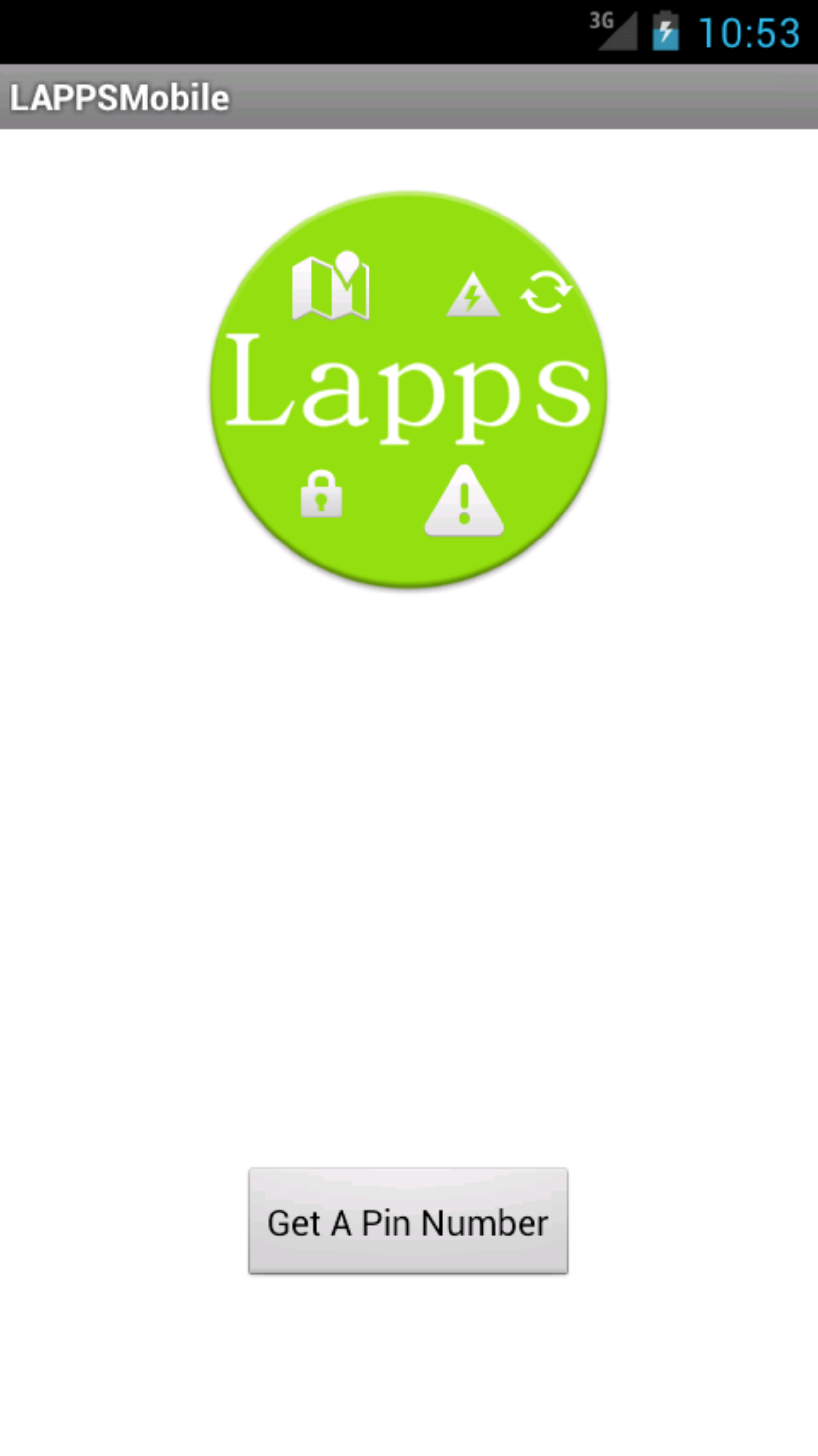}\label{fig:lapps_mobile_ss_a}}
    \subfloat[Application requests for a 8 digit number]{     \includegraphics[scale=0.228]{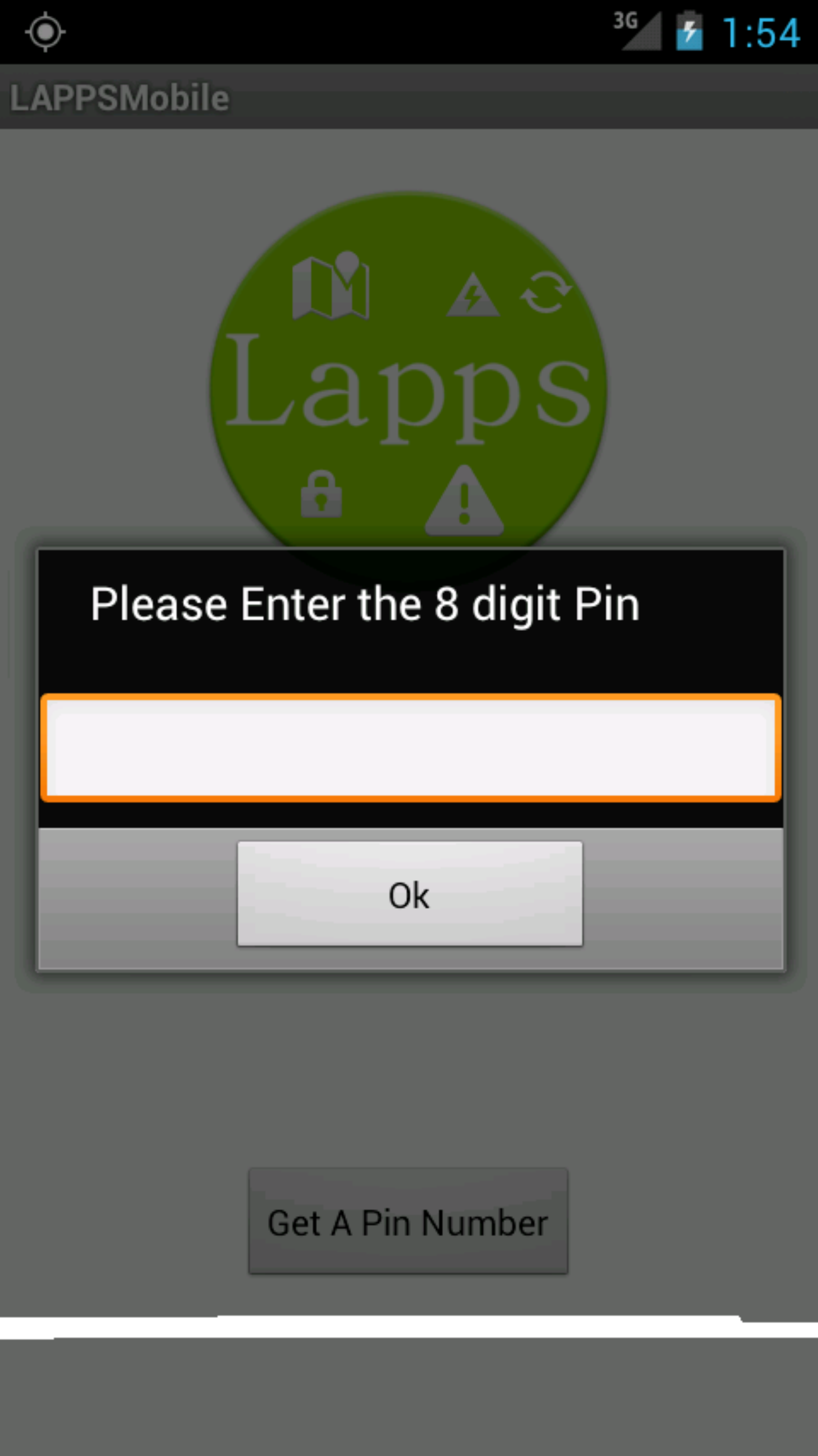}\label{fig:lapps_mobile_ss_b}}
    \subfloat[A Successfully received QR code]{    \includegraphics[scale=0.228]{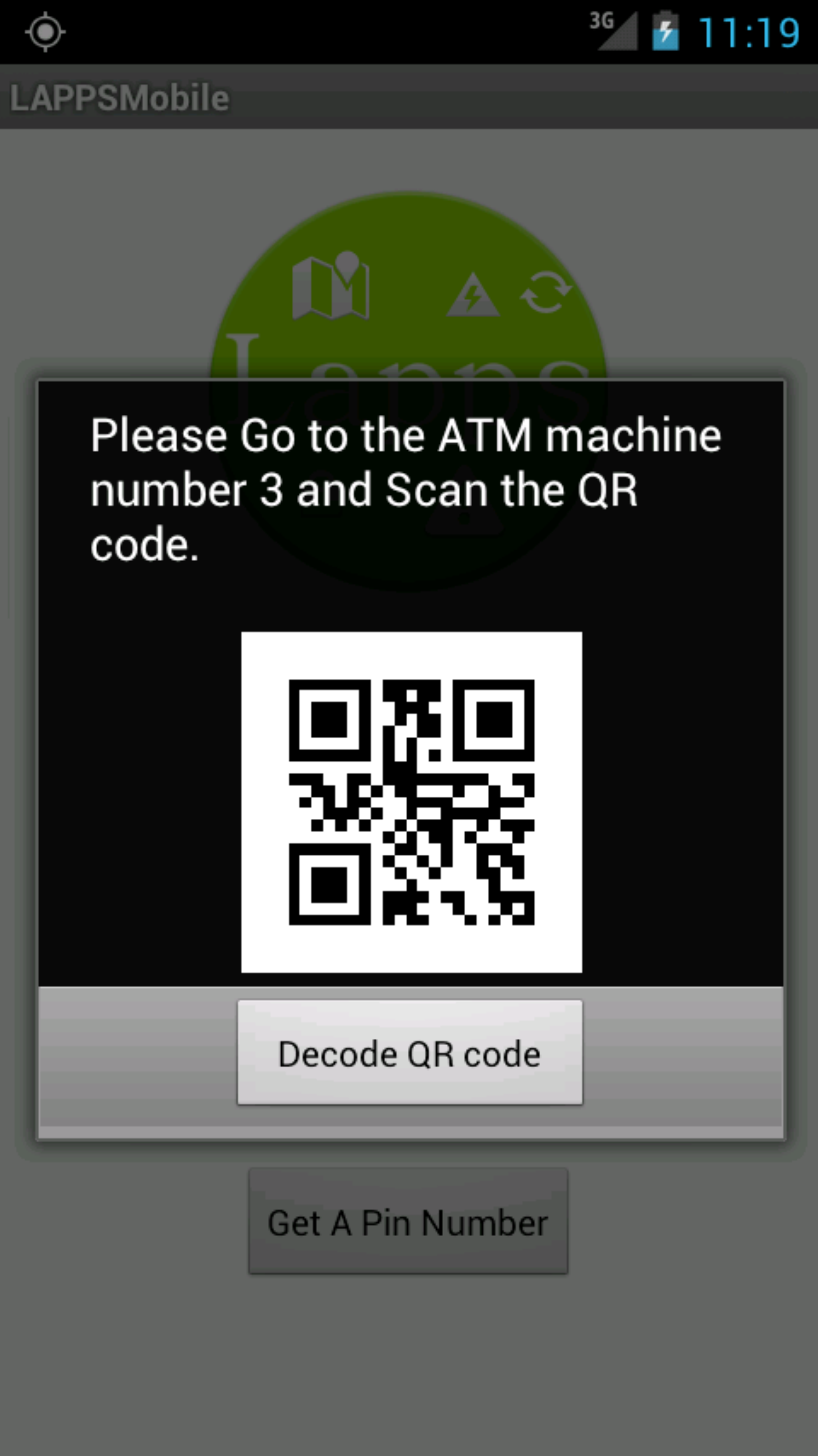}\label{fig:lapps_mobile_ss_c}}
    \subfloat[After the QRCode has been decoded]{
      \includegraphics[scale=0.228]{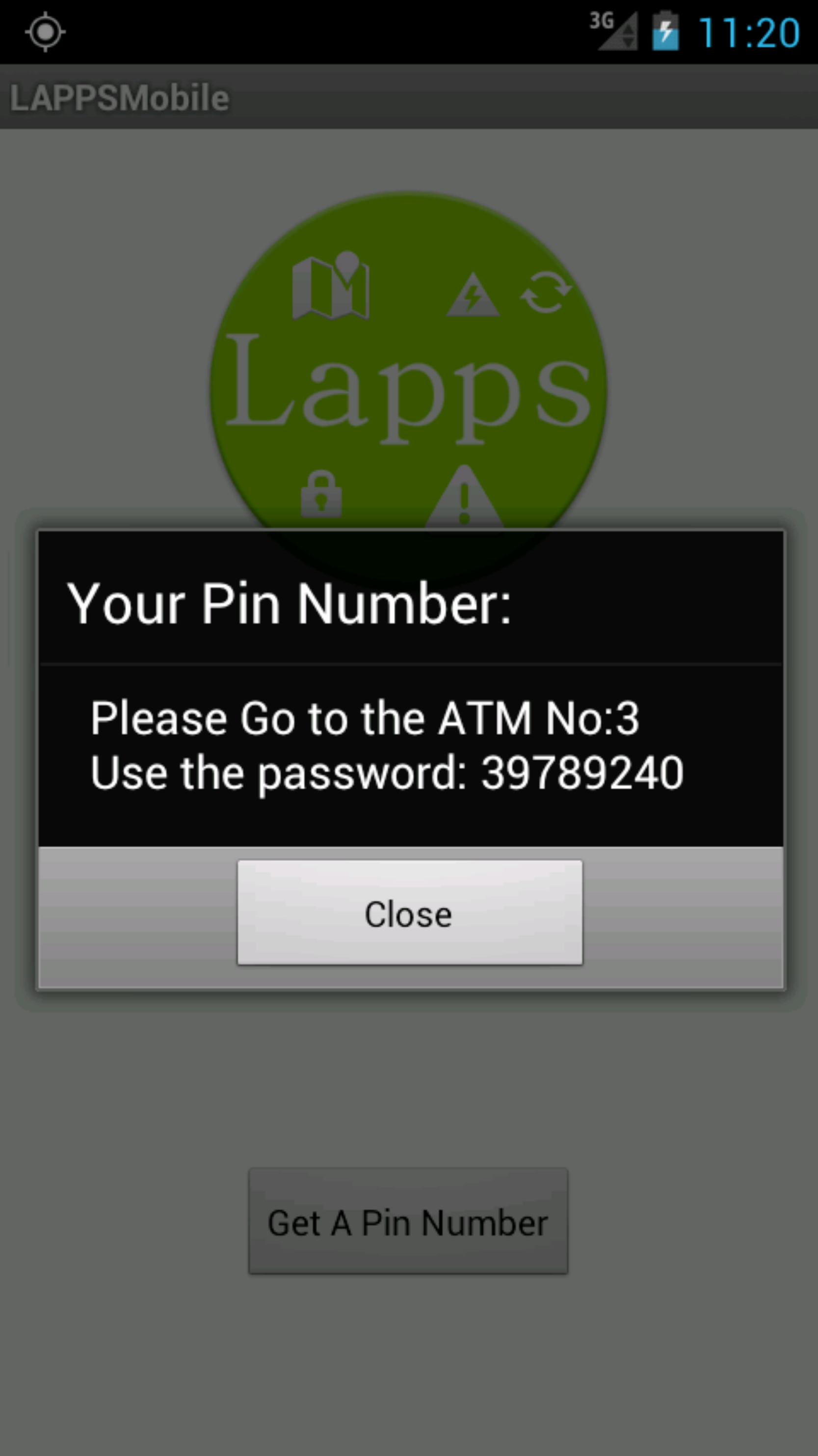}\label{fig:lapps_mobile_ss_d}}
  \end{center}
  \caption{The LAPPSMobile Application Screen-shots.}\label{fig:lapps_mobile_ss}
\end{figure}

\paragraph{Software Design}\label{sec:lappsmobile_software_design}:\\

A class diagram of the software design of the LAPPSMobile application
is shown below in figure \ref{fig:lapps_mobile_design}

\begin{figure}[htp]
  \begin{center}
    \includegraphics[scale=0.3]{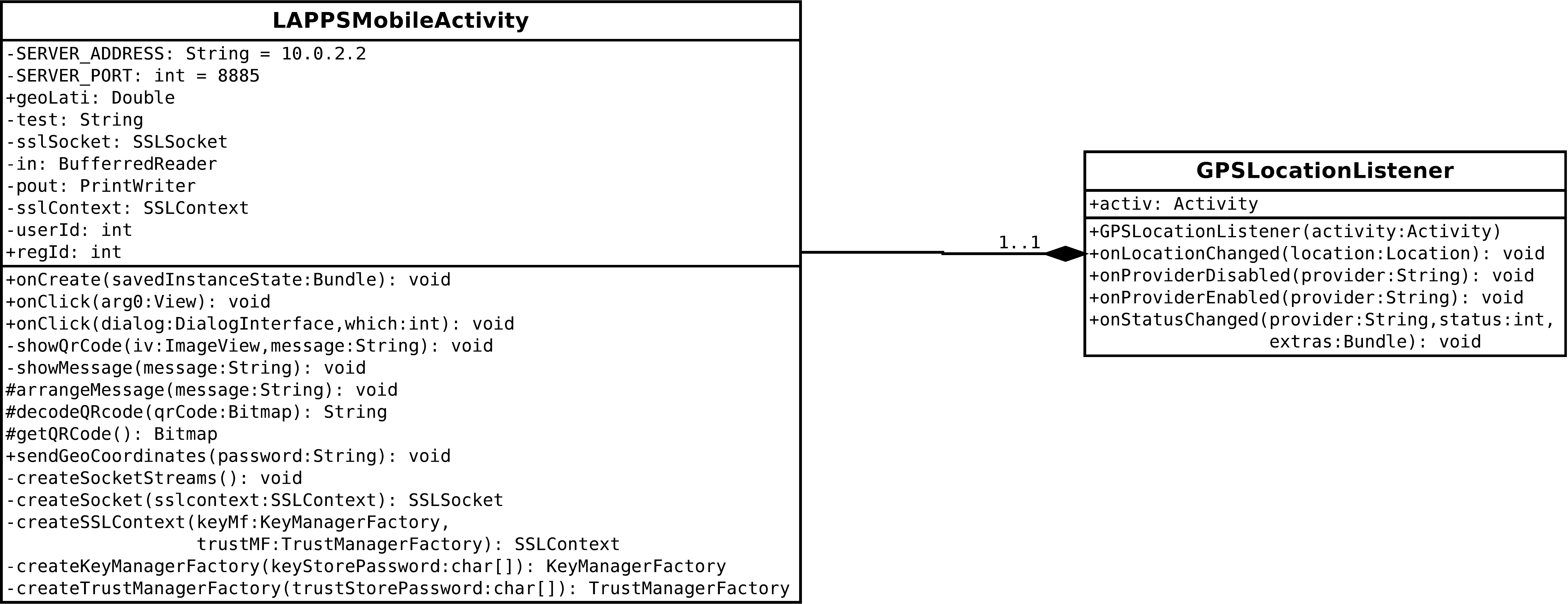}
  \end{center}
  \caption{The design of the LAPPSMobile Application}
  \label{fig:lapps_mobile_design}
\end{figure}

The application can be compiled and has been tested on android 2.2,
2.3.3, 3.2 and 4.03 platforms. The application seem to function very
well in all of these platforms.

The application requires following permissions to be able to get
access to the resources to be able to function substantially. The
permissions are programmed in XML language in the AndroidManifest.xml
file in the application.

\begin{verbatim}
<uses-permission android:name="android.permission.ACCESS_FINE_LOCATION" />
<uses-permission android:name="android.permission.ACCESS_COARSE_LOCATION" />
<uses-permission android:name="android.permission.INTERNET" />
<uses-permission android:name="android.permission.ACCESS_NETWORK_STATE" />
\end{verbatim}

The first line permits the access to the location information using the
GPS receiver of the mobile device. The Second line allows access to the
location information using Wifi, Cell-ID etc., to the mobile
application. Final two lines grant the access to the Internet and the
state of the network respectively.

\subparagraph{The work flow of the LAPPSMobile
  Application}\label{sec:lappsmobile_workflow}:\\

\begin{enumerate}
\item When the application starts, the LAPPSMobile application
  acquires a TLS over TCP/IP connection to the LAPPSServer. If the
  attempt to connect to the server is not successful a message is being
  displayed on the screen. When the user taps on the "Get a Pin
  Number" button (Figure \ref{fig:lapps_mobile_ss_a})  it will try to
  reconnect to the LAPPSServer in  the occasions that it is not connected.
\item When the user taps on the "Get A Pin Number" button the
  application prompts the user to enter an 8 digit number (Figure
  \ref{fig:lapps_mobile_ss_b}) only if the application can retrieve
  location information. If the GPS information is unobtainable then a
  message is being displayed on the screen (More on retrieving GPS
  information in section \ref{sec:gps_loaction_listener}).
\item After the user has tapped on the "Ok" button the location
  information, ID of the user, registered ID of the application and
  the inserted 8 digit number is sent to the server.
\item If the authentication is successful then the LAPPSServer sends a
  QR code with the allocated password and the ID of the \gls{atm}
  encoded.
\item The received QR code is displayed on the screen with the ID of
  the ATM machine that the user and the password have been allocated
  to (Figure \ref{fig:lapps_mobile_ss_c}). Ideally the user is able
  to scan the QR code on the correct \gls{atm} and log in to the
  machine. The second option is to tap on the "Decode QR code" button so that
  the QR code will be decoded (More on decoding QR code in section
  \ref{sec:qrcode_decode}) and the password and the ID of the
  \gls{atm} is presented in the next window (Figure
  \ref{fig:lapps_mobile_ss_d}).
\item Finally the connection closes when the user exits from the application.
\end{enumerate}

\subparagraph {How LAPPSMobile handle response messages}\label{lappsmobile_response_messages}:\\

There are currently two types of respond messages. They are "SUCCESS"
messages which are successful responds and "FAIL" messages which are
responds that are sent back for unsuccessful requests. Comparatively
to the way LAPPSServer has been coded to handle request messages (in
section \ref{sec:client_server_communication}) the LAPPSMobile has
been coded in consideration of that it is easily adaptable for future
changes as well as the clarity of the code itself. A pseudo code of
the code is shown below. To demonstrate how a newly introduced response
message (NEW\_RESPOND) can be handled a dummy code block has been added to the pseudo
code.

\begin{verbatim}
while (true) {
  if (connection.hasRespond()) {
   
/****Deals with "SUCCESS" requests***/ 
   if (connection.next().equals("SUCCESS:")) {
      displayQRCode();
   }
/**********************************/ 

/****Deals with "FAIL" response messages***/ 
   if (connection.next().equals("FAIL:")) {
      displayErrorMessage();
   }
/**********************************/ 


/****Deals with "NEW_RESPOND" response messages***/ 
   if (connection.next().equals("NEW_RESPOND:")) {
      handleResponse();
   }
/**********************************/ 

   // Code for handling new response codes can be added here
}
\end{verbatim}

\subparagraph{Decoding the QR codes} \label{sec:qrcode_decode}:\\\\
After the application has been received the QR code, LAPPSMobile uses
and instance of the "com.google.zxing.qrcode.QRCodeReader" class from
the "Zxing" library to decode the received QR code. The "Zxing" class
"RGBLuminanceSource" has been added supplementary to the existing
"com.google.zxing.qrcode" library. This file is licenced under the
Apache License, Version 2.0 \cite{apache_license2}. The
"RGBLuminanceSource" has been used as a "LuminanceSource" when converting the
"Bitmap" typed QR code to type of "BinaryBitmap".

\subparagraph{GPSLocationListener class}\label{sec:gps_loaction_listener}:\\

The LAPPSMobileActivity class contains an instance of the
GPSLocationListener class. This class is a type of "LocationListener"
which updates the user's location persistently and stores them in two
latitude and longitude variables in the "LAPPSMobileActivity"
class. Additionally it notifies the main application when the location
information is obtainable and unobtainable.

\subsubsection{ATM emulator}\label{sec:atm_emulator}

The LAPPSMobile is able to request for passwords and LAPPSServer is
capable of generating passwords. One of the important aspect of this
project is to authenticate and to prove that the generated passwords
ensures the security measures that LAPPS claims to comply.

\begin{figure}[htp]
  \begin{center}
    \includegraphics[scale=0.5]{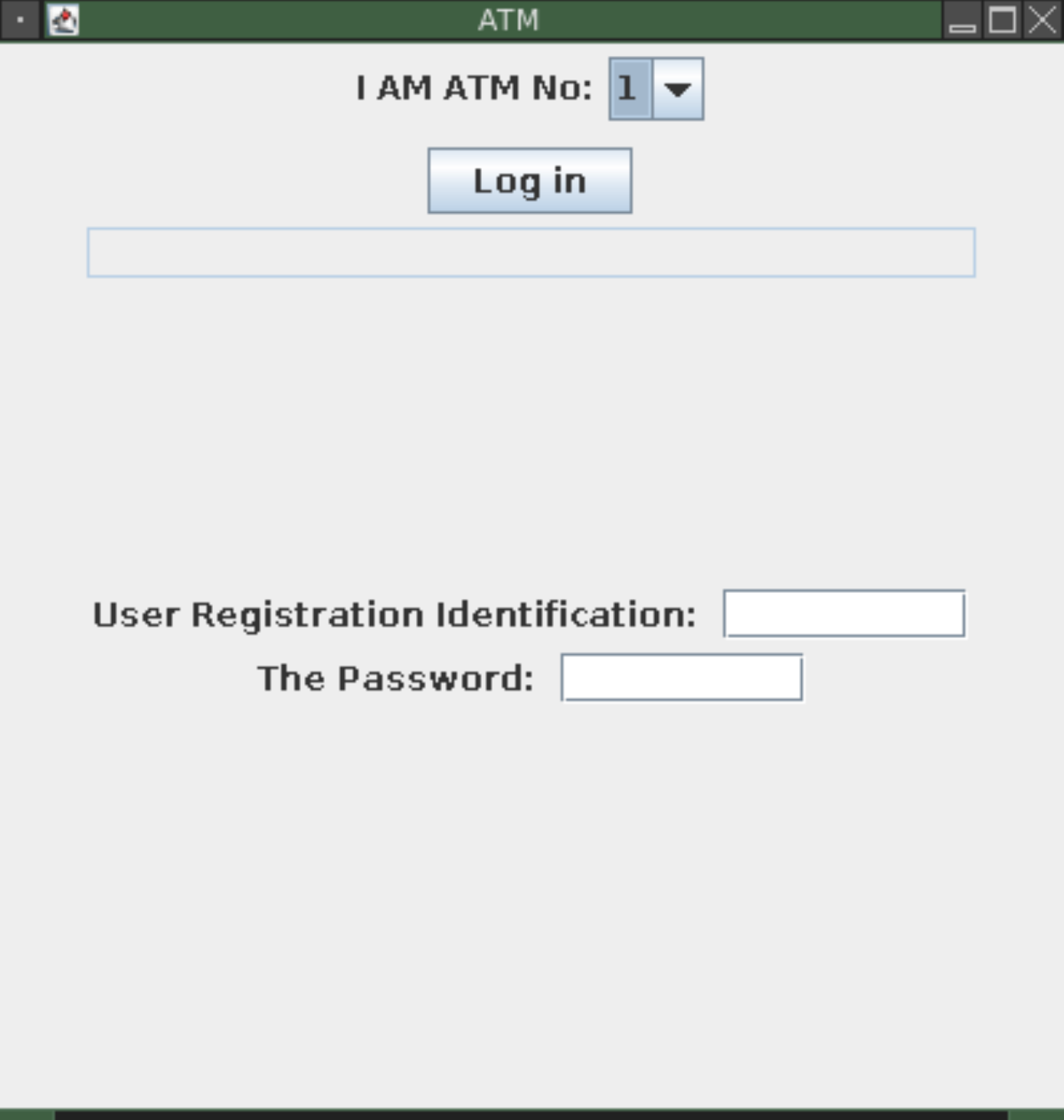}
  \end{center}
  \caption{Screen-shot of the ATM Emulator \label{fig:atm_emu_ss}}
\end{figure}

\paragraph{Software Design}\label{sec:atm_emulator_design}:\\

The following class diagram shows the design of the software program.

\begin{figure}[htp]
  \begin{center}
    \includegraphics[scale=0.4]{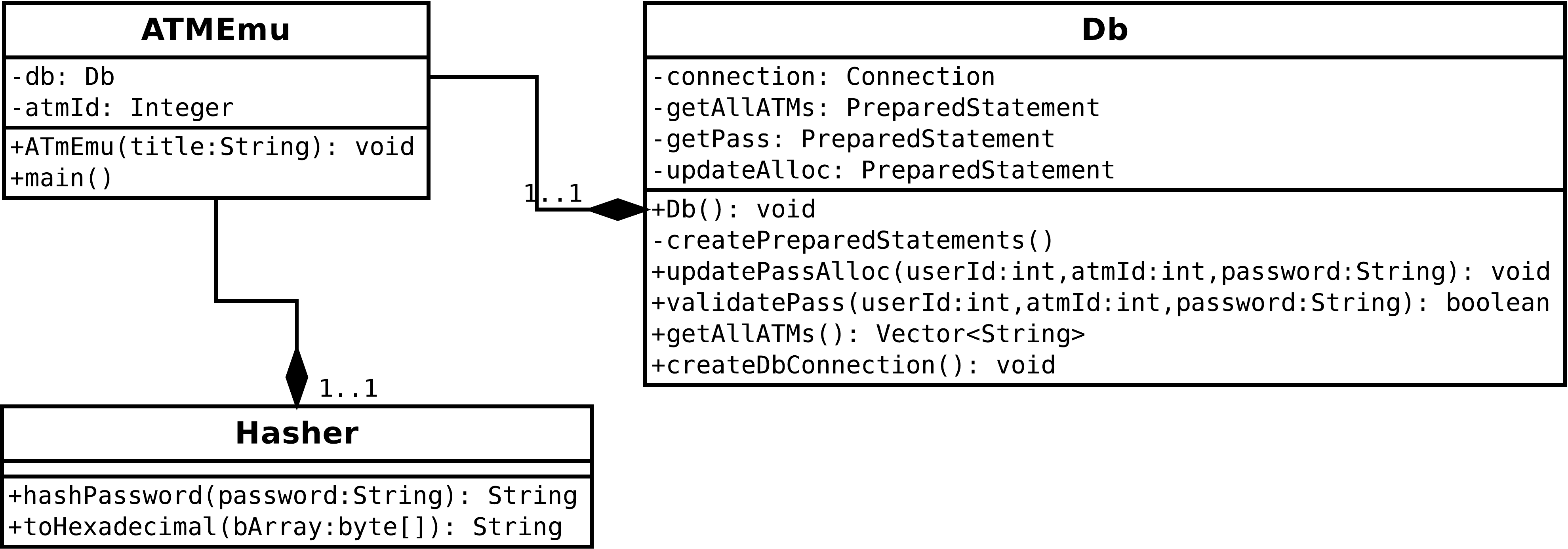}
  \end{center}
  \caption{The design of the ATM Emulator \label{fig:atm_emu_design}}
\end{figure}

The diagram above shows the attributes and the functions in the
classes and it suggests the relationships between them. ATMEmu class
contains instances of Hasher and Db classes (1..1). 

\subparagraph{How a user would use the ATM emulator}:\\

\begin{enumerate}
\item User selects the ATM machine that their password has been
  allocated to. This resembles the user using the allocated ATM
  machine to use the received password.
\item Then user enters their Identification number(ID). This resembles
  the user inserting their Credit/Debit card to the allocated \gls{atm}
\item Finally the user enters their fixed password and clicks on the
  "Log In" button.
\item If all of the authentication information is correct then a
  message is shown on the screen. Alternatively if the information is
  not correct an error message is shown.
\end{enumerate}

\subparagraph{The work flow of the ATM
  emulator}\label{sec:atm_emu_workflow}:

\begin{enumerate}
\item ATMEmu creates the Graphical User Interface when it starts.
\item After a user has entered the authentication information, it
  authenticates these information by querying the lappsStore
  database (Pseudo code of the query in section
  \ref{sec:atm_pas_auth}). Because the password has been stored in the
  database as hashed using SHA2-512, ATMEmu validates this password by
  creating a hash value of the inserted password and comparing it with the value that is stored in the database. To accomplish this the "hashPassword()" method in the Hasher class has been used (More on password authentication in
  section \ref{sec:atm_pas_auth}).
\item If the password is not authoritative then an error message is
  shown on the screen.
\item If the user has successfully logged in, ATMEmu stores the value
  "true" in the "used" column in the corresponding row in the
  "allocation" table. This indicates that the password has been used.
\end{enumerate}

\subparagraph{Authenticating Password}\label{sec:atm_pas_auth}:\\

To authenticate the password that is being inserted in to the ATM
Emulator, it queries the lappStore database to validate if there is a
predefined allocation to the particular \gls{atm} that belongs to the
particular user. Then the inserted password is matched against the
password that is stored in the database. Finally it checks if the password's
expiry time has been passed and it has not been used. If all of the above
parameters are positively ticks then one record will be returned.

The following pseudo code of a SQL query shows how this is done.

\begin{verbatim}
SELECT atmId 
FROM allocation AS a,password AS p 
WHERE a.atmId=<ATM_ID> AND a.userId=<USER's ID> AND
a.password=<Inserted_password AND a.password=p.password 
AND p.expTime > CURRENT_TIMESTAMP AND a.used=false;
\end{verbatim}

If the above SQL query returns a record then it is confirmed that the
inserted password is authoritative.

If the password is correct then it means that the user will log in to
the \gls{atm}. Hence it has to be noted immediately in to the
database. This is done by using a SQL query and it is demonstrated by the
following pseudo code.

\begin{verbatim}
UPDATE allocation SET used=true WHERE atmId=<ATM_ID 
AND userId=<USER's ID> AND password=<PASSWORD>
\end{verbatim}

\section{Performance of LAPPS}\label{sec:lapps_performance}
In this section the author looks at the performance of the LAPPS
Server. To carry out the test runs he has used an Intel(R) Core(TM)2
Duo CPU P8600 @2.40GHz CPU with L2 3072 KB of cache. The laptop has
got 4 GB of RAM and using Ubuntu Linux 11.10.

To send request to the server in this experiment he has used the
Android emulator that comes with the Android SDK \cite{android_sdk}, to run the
LAPPSMobile application. The server, Pin generating device and
the client (Android emulator) were running on the same machine.

Table \ref{tab:performance} shows the performance measures of LAPPS in
different test cases, that the author has gathered after performing 80 runs per
each test case. The median total response time of LAPPS is 56
milliseconds, despite the fact that all programs were running in the
same computer for testing so that the processing power has been distributed
between all programs. Table \ref{tab:performance} shows that the
median time that has taken to authenticate an 8 digit pin number was
about 1 millisecond and it is 1\% of its total response time. Similarly,
to find the nearest ATM to the user and to generate a new unique
password, 1 millisecond has been taken in both cases and both measured
times are 1\% of their total response times. To generate a QR code for
a response message, the median time that LAPPS has taken was 3
milliseconds. This is 5\% of its total response time. Finally time
wise the most expensive action in the process of responding to user
requests was storing allocation records in to the database. In median
time, it has taken 39 milliseconds, that is 69\% of its total response time. 

We can conclude that the current LAPPS implementation is fairy quick
although the performance evaluation could have carried out in a better
environment for better results. This can be added as a further task on
this project.

\section{Evaluation}\label{sec:evaluation}

The Location Aware Password Protection system is an architecture that
adds five assorted layers around passwords that is generated by
traditional password protection systems. These layers have been
designed to harden the traditional password protection systems. In
section \ref{sec:lapps_arch} the author has expounded how these layers
work. The main advantage of these layers is that according to the
user's desire, the layers can be added or removed. The main speciality of the
whole \gls{lapps} architecture is that the passwords can be allocated to
geographical locations, so that a password will only be valid in it's
allocated geographical area. Also the password can only be allocated to
one geographical area, even though you can choose to customise the
"Location Awareness" layer so that you can allocate the password in to
varied geographical areas. Though these layers harden the passwords
there still are drawbacks. These are discussed in section
\ref{lapps_eval}. LAPPS allows one password only to be used once, by
storing a Boolean value in the 'used' column in the 'allocation'
table, if one has been used. A password's active time has been
restricted by storing the expiry time in the database, so that
only if the current time is less than the expiry time, the password is
usable. The Pin Sentry device introduces a second authentication
factor that hardens the security.

The current implementation of \gls{lapps} has been implemented to harden the
password security systems in Automated Tailor machines. Though it has not
been tested on a real \gls{atm} network yet. Section
\ref{sec:layer_mapping} demonstrates how the \gls{lapps} layers have 
directly been implemented into a real world software
application. Section \ref{sec:components} goes through more technical details on how these components of "\gls{lapps} for \glspl{atm}" have been
implemented.

\section{Project Planning}\label{sec:project_planning}

I have started planning my project very early. As early as June 2011; just after I have finished the second year I have had a meeting with
my supervisor regarding my allocated project "Live video stream over
cloud". After a couple of discussions I was given the opportunity to
deliver my own project proposal that is relative to my interest
areas. Since "Computer Security" is one of my interest areas, I have
proposed a project idea and we named it "Location Aware Password
Protection System".

Through out the project planning, designing and the implementation
processes I have regularly noted the changes, new findings,
achievements and etc. in my logbook, and the finished milestones and task had been marked in the gantt chart. The log book has  been regularly marked by
Prof.~Kun after every meeting that we did once every two weeks.

Since the day I came up with the \gls{lapps} idea (07/09/2011), I have
researched about the theoretical values of the technologies that is
relevant to the subject area. This helped me to gain a better
understanding of the topic. At first the project scenario was just based on
the security of \glspl{atm}. After 19/09/2011 I have started looking in to
more technical details that could be imperative for the implementation
of the project, as well as the more theoretical sides. The background
research gained me more confidence about the plausibility of
implementation of the project that has to be done in a limited time
period. At this stage I was quite confident about the project idea and
the implementation, and I had planned to finish off the project during the December holiday in 2011/2012. 

On 27/11/2011 I started the design of the software with a draft
design of the database. The initial idea of the project has been
subject to continuous minor changes over time. For instance, according
to the newly evolved project idea, the password that will be generated
by \gls{lapps} had to have a time frame that it will be valid
for. Consequently the design of the database had to be changed. This
was done on 03/01/2012. There were three main components in the
software system and by this time I had already started the development
of all of the components together. Since all of the components have to
work together, testing had been carried out along side with the
software development. I was quite happy that I stared the
implementation very early so I would be able to get on with other
university work with less pressure when the university starts. The
changes that my supervisor proposed and the ones that I have decided
to change, I happily accepted and put in to the design because they
only made the project better and more interesting.

To keep the project on track and manage the code, I have used the
"git" version control system. This has made my work easier to manage
and to keep track of. Every time I have applied a significant change to
the code, I have committed the changes to the "git" local
repository. Additionally I submitted comments along with the commits
so I have a description of what I have done corresponding to the
commits. The daily progress had been noted on the log book. The
research I have been doing during the summer and the first semester helped
to choose the technologies and tools to implement the software.

During the implementation, there were easier problems than some of the
others and there were more difficult problems than many others. It is
fair to say that this project is fairly a complicated one. One of the
challenges that I had to face during the development was that I had
problems with implementing TLS mechanism to the socket connections in
the Android application. I only had the problem when I compiled this
code for the Android 2.2 framework. As a solution I carried on my
development under the Android 4.03 framework. Hence I was able to
carry out the development of the project until I had found an alternative
to the bug I had under the Android 2.2 framework. I was able to
complete the implementations of all of the components before the first
meeting with my supervisor on 26/01/2012. The effort payed off at the end because I had a good feed back about the finished project, from Prof.~Kun.

Having completed the proposed password protection system, after few
discussions with my supervisor I decided to extend the project
considering that I had plenty of time left. After the first extension
the passwords that are stored in the database are no longer stored as
plain text, now database only holds the hashed values of the
passwords, consequently I had to program ATM Simulator and the
LAPPSServer to work with the hashes. Secondly I have introduced a Pin
generating device software to the software system. So that all users
are required to supply the 8 digit number that is generated by this
device software when they request for a password. Consequently I have changed
LAPPSMobile Android application, the database and the LAPPSServer
application to work with the 8 digit pin number and to authenticate these digits. All of the these had been completely implemented by
22/02/2012. Figure \ref{fig:gantt1} and \ref{fig:gantt2} shows the
project task breakdown and gantt charts.

Looking back at all of the phases in the project, I think I did well
in terms of time management of the project. Hence I had extra time to
extend, test and modify the software among many other assignments and
coursework at the University. There were many changes that I have
carried out during the implementation such as I have decided to change
the Database Management System (DBMS) to PostgreSQL so that I was able
to use the PostGIS software program. So I had to adapt to its SQL
syntax for queries and triggers. The whole process of learning Android
development feels like an adaptation from general Java programming to
Android. Although I was enjoying every bit of it. In the project I
have learnt so many things such as Project Planing, about TLS/SSL
keys and certificate generation, Android application development,
PostgreSQL DBMS and git version control system. I believe with the
experience I have gained during this project I will be able to do a
better job next time I will be involved in a project.

\section{Summary}

In this thesis the author has introduced an architecture that is designed to
consolidate conventional password protection systems. \gls{lapps}
architecture can be adopted to the traditional password protection
systems easily without having to comprehensively replace old
systems. These layers make the traditional passwords safe enough to
protect private and confidential information considering the threats
that have been reported for the last decade. Additionally these layers are
customisable to the user's needs. The signature value of \gls{lapps} is the
location awareness.  

The "\gls{lapps} for \glspl{atm}" implementation is an example of how this
architecture can be endorsed into a software application. Section \ref{sec:evaluation} shows how the objectives
\ref{sec:objectives} of the project have been achieved. We have seen
the performance of LAPPS in different test cases in section
\ref{sec:lapps_performance}. Finally section \ref{sec:project_planning}
discloses how the author has manged the time during the project.

\bibliographystyle{IEEEannot}
\bibliography{citations}

\appendix
\section{Appendix}

\begin{figure}[htp]
  \centering
  \begin{center}
    \includegraphics[scale=0.37]{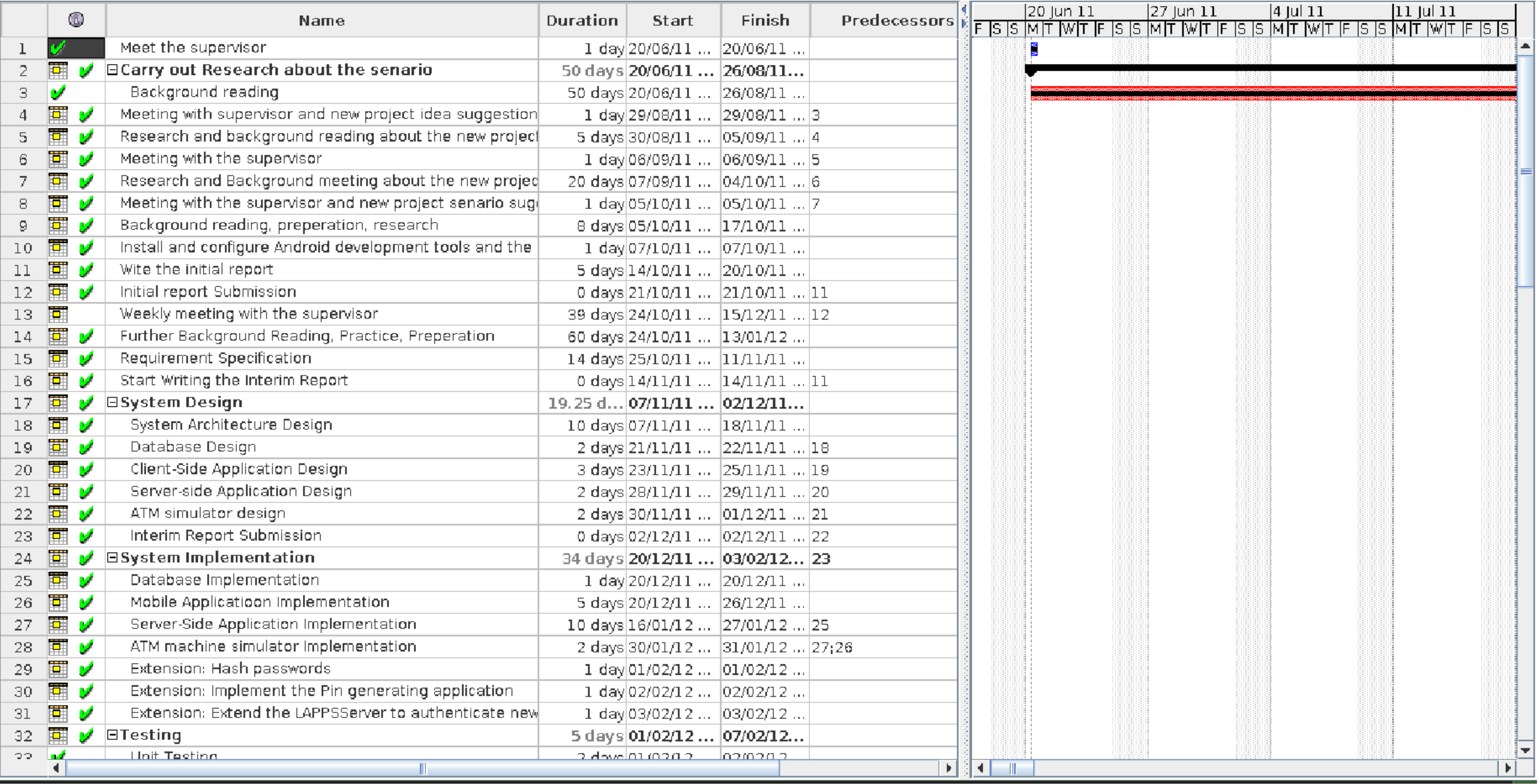}
  \end{center}
  \caption{Task breakdown and gantt Chart 1 \label{fig:gantt1}}
\end{figure}

\begin{figure}[htp]
  \begin{center}
    \includegraphics[scale=0.37]{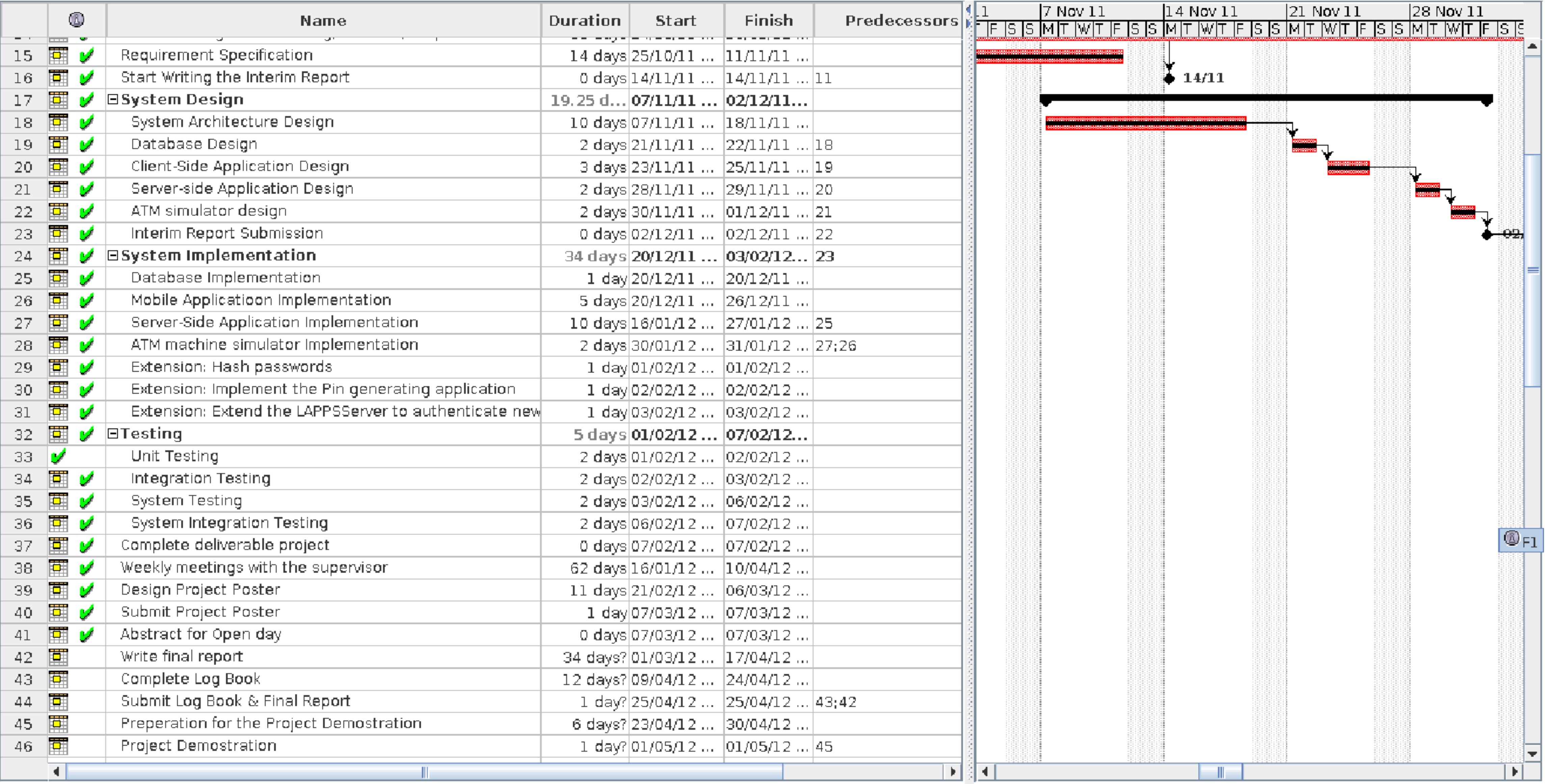}
  \end{center}
  \caption{Task breakdown and gantt Chart 2 \label{fig:gantt2}}
\end{figure}

\begin{table}
  \begin{tabular}{|l|c|r|}
    \hline

    Test Case & Median Time (ms)& \% from total response time\\
    \hline
    Total response time & 56 & 100 \\
    Pin number authentication & 1 & 1\\
    Generate unique password & 1 & 1\\
    Find closest ATM & 1 & 1 \\
    QR code generation & 3 & 5\\
    Store allocation & 39 & 69\\
    \hline
  \end{tabular}
  \caption{The performance measures of LAPPS}
  \label{tab:performance}
\end{table}
\end{document}